\documentclass[aps,prd,preprint,superscriptaddress,preprintnumbers,nofootinbib]{revtex4-1}
\pdfoutput=1

\usepackage{graphics}

\usepackage[dvips]{graphicx}
\usepackage{mathrsfs}
\usepackage{amssymb}
\usepackage{amsmath}
\usepackage{amsfonts}
\usepackage{verbatim}
\usepackage{float}
\usepackage{placeins}
\usepackage{slashed}
\usepackage{bbm}
\usepackage[dvips,letterpaper,text={6.5in,9in}]{geometry}

\usepackage[english]{babel}
\usepackage{tabu, multirow,color,dcolumn,bm,enumerate}

\newcommand{\tev}{\text{TeV}}
\newcommand{\mev}{\text{MeV}}
\newcommand{\gev}{\text{GeV}}

\usepackage{graphicx,bm}

\newcommand{\beq}{\begin{equation}}
\newcommand{\bea}{\begin{eqnarray}}
\newcommand{\eeq}{\end{equation}}
\newcommand{\eea}{\end{eqnarray}}
\newcommand{\bal}{\begin{align}}
\newcommand{\eal}{\end{align}}

\usepackage{tikz}
\usetikzlibrary{decorations.pathmorphing,decorations.markings}
\tikzset{
photon/.style={decorate, decoration={snake,amplitude=4pt, segment length=7pt}, draw=black},
particle/.style={draw=black, postaction={decorate}, decoration={markings,mark=at position .5 with {\arrow[draw=black]{>}}}},
antiparticle/.style={draw=black, postaction={decorate}, decoration={markings,mark=at position .5 with {\arrow[draw=black]{<}}}},
gluon/.style={decorate, draw=black, decoration={coil,amplitude=3pt, segment length=4pt}},
higgs/.style={draw=black,dashed,thick },
arrow/.style={draw=black, very thick, postaction={decorate}, decoration={markings,mark=at position 1 with {\arrow[draw=black]{>}}}}
}

\usepackage{latexsym}
\usepackage{mathrsfs}
\usepackage{amsthm}

\definecolor{darklightsabergreen}{rgb}{0.0, .49, 0.06}

\begin{document}

\title{Constraints on  $L_\mu - L_\tau$ interactions at the LHC and beyond}

\author{Fatemeh Elahi}
\affiliation{Department of Physics, 225 Nieuwland Science Hall, University of Notre Dame, Notre Dame, IN 46556, USA}
\author{Adam Martin}
\affiliation{Department of Physics, 225 Nieuwland Science Hall, University of Notre Dame, Notre Dame, IN 46556, USA}

\vspace*{0.5cm}
\begin{abstract}
\vspace*{0.5cm}

{In this paper we examine the constraints dedicated LHC multi lepton searches can place on $Z'$ bosons coming from gauged muon number minus tau number, $L_{\mu}-L_{\tau}$. As the  $L_{\mu}-L_{\tau}$ gauge boson does not couple to proton constituents or electrons at tree level, the current bounds are fairly loose, especially for $M_{Z'} \gtrsim 1\, \gev$. For $2m_{\mu} < M_{Z'} < M_Z/2$ we develop search strategies using the $pp \to Z \to 4\, \mu$ channel. The cleanliness of the final state, combined with the fact that $pp \to Z \to 4 e$, $Z \to 2e\,2\mu$ can be used as background control samples, allow us to spot $L_{\mu}-L_{\tau}$ $Z'$ with couplings $\mathcal O(10^{-3})$ times the Standard Model couplings. For lighter $Z'$, we propose the mode $pp \to 2\mu + \slashed E_T$. The presence of missing energy means there is a wider set of backgrounds to consider in this final state, such as Drell-Yan production of leptonically decaying $\tau$ pairs, however we find these can be controlled with careful cuts. Combining the $4\, \mu$ and $2\,\mu + \slashed E_T$ modes, we find that with $\sim3 \, \text{ab}^{-1}$ of integrated luminosity we are sensitive to couplings $g_{Z'} \gtrsim 0.005\, g_1$ and $0.5\, \gev \le M_{Z'} \le 40\, \rm \gev$ and $g_{Z'} \gtrsim 0.001\, g_1$ for $M_{Z'} < 2\, m_{\mu}$. This region includes the parameter space where  $L_{\mu}-L_{\tau}$ models can ameliorate the muon $g-2$ anomaly. We repeat these analyses at a future $e^+e^-$ Z-factory, where we find improved sensitivity. Specifically, given $2.6\, \rm ab^{-1}$ of luminosity, we can exclude $g_{Z'} \gtrsim 0.001\, g_1$ for $2\, m_{\mu} \le M_{Z'} \le M_Z/2$. } 
\end{abstract}
\maketitle

\section{ Introduction}
\label{sec:Intro}

With the discovery of the Higgs boson, the last piece of Standard Model (SM) is in place, and the next run of the Large Hadron Collider (LHC) will be dedicated to the search for traces of physics Beyond the Standard Model (BSM). Many extensions of the SM predict a new massive gauge boson, generically called $Z'$, that is electrically neutral, and a color singlet. Among the proposed candidates, the massive $Z'$ formed from gauging the difference between muon lepton number and tau lepton number,  $L_\mu-L_\tau$ is an interesting candidate to look for at LHC

The $L_\mu-L_\tau$ $Z'$ is particularly interesting to hunt for at the LHC because the current constraints are not nearly as strong as other $Z'$. Any $Z'$ that couples significantly to light quarks is severely constrained, and the bounds only fade off for very heavy $Z'$ ( $M_{Z'} \sim O( TeV)$ with coupling of $O(1)$) \cite{Schuh:2015hda,Charaf:2015jua}. One way to avoid these collider constraints is to have a $Z'$ that does not couple at tree level to quarks, a `hadrophobic' $Z'$, and a simple way to arrange this is to promote lepton number to a gauged symmetry. Lepton number, by itself, is anomalous if we include only SM matter field content, and thus cannot be gauged. However, the differences between the lepton numbers of different generation leptons, $L_e - L_\mu$, $ L_e - L_\tau$, and $L_\mu - L_\tau$ are anomaly free \cite{ He:1991qd}. 

Light $Z'$ that couple to electrons have been severely constrained. For $M_{Z'} < 10 ~ \gev$, BaBar has been able to limit a new gauge boson coupling to electron at level of $O(10^{-3})$ \cite{Eigen:2015rea,Curtin:2014cca,Eigen:2015sea}, and the projected exclusion reach from Belle II is an order of magnitude stronger~\cite{Essig:2013vha,Wang:2015hdf}. Due to these strong low-energy constraints, we will focus here on the lepton number combination that does not involve electrons, $L_\mu -L_\tau$.  A further reason to study $L_\mu - L_\tau$ is the persistent discrepancy between SM prediction and experimental measurement of muon anomalous magnetic moment: $(g-2)_{\mu} $ \cite{Anastasi:2015oea}. Although this anomaly may be due to theoretical uncertainties from QCD contributions, a $Z'$ that only interacts with second and third generation of leptons can also explain the discrepancy~\cite{Baek:2001kca,Ma:2001md,Gninenko:2001hx,Pospelov:2008zw,Heeck:2011wj,Harigaya:2013twa} in certain regions of parameter space. Gagued $L_\mu-L_\tau$ can also explain some of the observed anomalies in B physics and flavor changing Higgs coupling as discussed in \cite{Crivellin:2015lwa,Crivellin:2015mga,Heeck:2014qea}.

One of the few constraints on the $L_\mu - L_\tau~Z'$comes from the CCFR experiment~\cite{Mishra:1991ws, Mishra:1991bv}, based on the process  $\nu_\mu + N \rightarrow \nu_\mu + N + \mu \mu$. These bounds are strongest for $M_{Z'}\, \sim\, \gev$  \cite{Geiregat:1990gz,Mishra:1991ws,Altmannshofer:2014cfa}. Above that mass, the authors of Ref.~\cite{Altmannshofer:2014cfa} found that stronger bounds on $L_\mu - L_\tau$ $Z'$ could be derived from the LHC run-I measurement of the rare process $pp \to Z \to 4\mu$. In the SM, $Z \to 4\mu$ comes about via $Z \to \mu^+\mu^-$, where one of the muons radiates a $Z^*/\gamma^*$ that creates the additional muon pair. In the presence of a $L_\mu - L_\tau$ $Z'$, there is an additional way for one of the muons connected to the $Z$ boson to radiate the extra pair of muons. Provided the $Z'$ is on-shell, it can impact the $4\,\mu$ rate and kinematic distributions even if the $Z'-\mu-\mu$ coupling is very weak.

The LHC bound discussed in Ref.~\cite{Altmannshofer:2014cfa} was derived using the default ATLAS  $4$-lepton analysis and was not optimized to pick out the kinematic traces of a $Z'$. The goal of this paper is to carry out this optimization and determine how well a high luminosity LHC (HL-LHC) run or future $e^+e^-$ Z-factory can constrain $L_\mu - L_\tau$. For $M_{Z'} > 2\, m_{\mu}$, we stick with the same channel as Ref.~\cite{Altmannshofer:2014cfa}, $pp \to Z \to 4\,\mu$. For $M_{Z'} < 2m_\mu$, $Z'$ can no longer go on-shell in $4\mu$ events, so we propose a different mode, $p p\rightarrow 2 \mu + \slashed E_T$. Some new backgrounds emerge in this channel, however they are kinematically very different from our signal and thus they can be removed.\footnote{The proposed experiment at CERN \cite{Gninenko:2014pea} will also be sensitive to the region of parameter space we are considering. LHC constraints for heavier $Z'$ is discussed in \cite{delAguila:2014soa}.}

The setup of the rest of the paper is as follows. In the next section, we introduce the model and discuss the existing constraints on its parameters based on precision measurements and work in Ref.~\cite{Altmannshofer:2014cfa}. In Sec.\,\ref{sec:analysis} we explore how the bounds can be improved,  both by extrapolating the existing searches to higher luminosity (\ref{sec:pp_fourmu}) and with optimized analyses~\ref{sec:optimal}. In Sec.~\ref{sec:pp_twomumet} we study $p p \rightarrow \mu^+ \mu^- ~ \slashed E_T$, which is better for $Z'$ lighter than twice the muon mass, then turn to $L_\mu - L_\tau$ studies at a future $e^+e^-$ Z-factory in Sec.~\ref{sec:ee_fourmu}. Finally a discussion about the results and concluding remarks are made in Sec.~\ref{sec:discuss}.


\section{ $L_\mu - L_\tau$ model}
\label{sec:model}

The SM possesses several accidental global symmetries. These global symmetries are anomalous, however one can form linear combinations of the symmetries that are anomaly free and can therefore be gauged. The $L_\mu - L_\tau$ model comes from exactly this concept, and the symmetry that is gauged is the difference between muon number and tau number~\cite{He:1991qd}. As we know, exact $L_\mu - L_\tau$ is not realized in nature,  $L_\mu - L_\tau$ needs to be broken.\footnote{This $U(1)$ must be broken, i.e. via the Stueckelberg mechanism, but the details of how it is broken are unimportant for our purposes. } As a result of this breaking, we have a massive,  neutral, color-singlet gauge boson. The parameters of the model are the $Z'$ mass $M_{Z'}$ and the $L_\mu - L_\tau$ gauge coupling. The Lagrangian is shown below:
\begin{align}
\mathcal L = \mathcal{L}_{\rm SM } - \frac{1}{4} (Z')_{\alpha \beta} (Z')^{\alpha \beta} + \frac{1}{2} M_{Z'}^2 Z^{'\alpha}Z'_{\alpha}  - \epsilon g_1 Z'_{\alpha} \left( \bar \ell_2 \gamma^\alpha \ell_2 + \bar \mu \gamma^\alpha \mu - \bar \ell_3 \gamma^\alpha \ell_3 - \bar \tau \gamma^\alpha \tau\right),
\end{align}
where $Z'_{\alpha \beta} = \partial _\alpha Z'_\beta - \partial_\beta Z'_\alpha$ is the field tensor, and $\ell_2 = (\nu_\mu, \mu)^T$, $\ell_3 = (\nu_\tau, \tau)^T$. In this model, the $Z'$ has the same coupling to left handed muon (tau) and right handed muon (tau), but there is a relative sign difference between the coupling of muons and taus~\cite{Altmannshofer:2014pba}. Notice that we define the $L_\mu - L_\tau$ coupling as a multiplicative factor $\epsilon$ times the hypercharge coupling. This is somewhat unconventional, but it allows us to easily estimate the size of $L_\mu - L_\tau$ effects. Working with this convention, our parameters are $\epsilon$ and $M_{Z'}$. In this setup, the width of  $Z'$ is narrow; the exact number of open decay modes depends on $M_{Z'}$, but all partial widths are proportional to $\epsilon^2\, g^2_1$.

Although the $Z'$ only couples to second and third generation leptons at tree level, loops of $\mu$ and $\tau$ induce $Z-Z'$ mixing. Redefining fields to remove this mixing generates a coupling between all fermions and the $Z'$ of $\mathcal O(g^2\, \epsilon\, \tan^2{\theta}\, \log{(m_{\tau}/m_{\mu}})/(48\pi^2)) \sim 10^{-3}\, \epsilon$. These loop corrections turn the $L_\mu - L_\tau$ setup into a generic dark photon setup for all fermions except the $\mu, \nu_{\mu}$ and $\tau, \nu_\tau$. For $10\, \mev \lesssim M_{Z'} \lesssim 10 \,\gev$, dark photon scenarios are tightly constrained by BaBar to have $Z'-$electron couplings $g_{Z'-e} \lesssim 10^{-3}$ times the electromagnetic coupling $e_{em}$~\cite{Curtin:2014cca}, which translates to a constraint of roughly $\epsilon \lesssim 1$ for the  $L_\mu - L_\tau$ model. For lighter $Z'$, the constraints are much stronger (coming from fixed-target experiments), while for $M_{Z'}' > 10\, \gev$ the current constraints are approximately $g_{Z'-e} \lesssim 10^{-2}\, e_{em}$ (corresponding to $\epsilon = 10$). The limits on $M_{Z'} > 10\,\gev$  have been projected to reach $g_{Z'-e}\, \lesssim 10^{-3}\, e_{em}$ ($\epsilon = 1$) by the end of the high luminosity LHC run~\cite{Curtin:2014cca}.\footnote{Kinetic $Z-Z'$ mixing also induces mass mixing between $Z$ and $Z'$, however this enters at second order in the kinetic mixing parameter, or $\sim 10^{-6}\, \epsilon^2$ in our setup. For the range of $\epsilon$ we are interested in, this shift is too small to provide any constraint.}

A surprisingly effective way to bound light $L_\mu - L_\tau$ scenarios with light $Z'$ is through the $Z'$-neutrino interaction. Specifically, neutrino beam experiments CHARMII collaboration~\cite{Geiregat:1990gz}, and CCFR collaboration~\cite{Mishra:1991ws, Mishra:1991bv}, have been shown~\cite{Altmannshofer:2014cfa} to be sensitive to $L_\mu - L_\tau$ $Z'$ via the `trident' process, muon pairs produced via scattering a muon-neutrino off of a nucleus: $ \nu_\mu + N \rightarrow \nu_\mu + N + \mu \mu $. In the SM, this process occurs through the exchange of $W^\pm/Z$ boson. As CHARMII and CCFR are fixed target experiments with incident beam energies of 23 GeV (CHARMII)  \cite{Beyer:1993jt} or $140\, \gev$ (CCFR)\cite{Smith:1992iw}, $\sqrt{\hat s} \ll M_W, M_Z$, so the intermediate bosons are always off-shell and the SM rate is small. This opens up sensitivity to trident production via the exchange of light, on-shell $Z'$. The fact that a light $Z'$ can go on-shell is crucial. An intermediate off-shell $Z'$ is suppressed by two powers of $\epsilon$ in the amplitude, one power at the production vertex and one at the destruction. On-shell exchange, on the other hand, only comes with the $\epsilon$ factor at the production vertex. The decay of the $Z'$ costs some $\mathcal O(1)$ branching ratio, rather than an additional $\epsilon$ factor. The study in Ref.~\cite{Altmannshofer:2014cfa} found trident production could bound  $L_\mu - L_\tau$ down to $\epsilon \sim 0.01$ for $M_{Z'} \sim \gev$. The bounds loosen as $M_{Z'}$ increases, rising to $\epsilon \ge 0.04$ at $M_{Z'} = 15\, \gev$ and $\epsilon \ge 0.1$ at $M_{Z'} = 30\, \gev$.  

For heavier $M_{Z'}$, the trident bounds are surpassed by bounds from the LHC.  While there are no dedicated searches for $L_\mu - L_\tau$ model at the LHC, bounds for this model may be derived from recasting other searches. One appealing channel to investigate is $pp \rightarrow Z \to 4\mu$ \cite{Altmannshofer:2014cfa}. In the SM, this final state is part of the higher-order correction to the di-muon decay $Z \to \mu^+\mu^-$, with one of the leptons radiating a $Z^*/\gamma^*$ that subsequently produces two additional muons. In the $L_\mu - L_\tau$ scenario, the initial muons can also radiate a $Z' \to \mu^+\mu^-$, and for $M_{Z'} \lesssim M_Z/2$ this $Z'$ can go on-shell.  Just as in the trident process, amplitudes with on-shell $Z'$ are only suppressed by one power $\epsilon$ (in the amplitude) and will be sensitive to weaker couplings than processes with off-shell $Z'$. Because of phase space considerations, heavier $Z'$ are more difficult to create on-shell than lighter $Z'$, and the highest mass we can probe in this channel is $M_Z/2$. There are several other benefits of looking at $pp \to Z \to 4\,\mu$: the final state is exceptionally clean, has no hadronic activity, and is difficult to fake. Additionally, by forcing the 4 muons to reconstruct an on-shell $Z$, essentially all background from continuum multi-lepton production is eliminated ($\lesssim 1\,\%$).

	Recasting the $pp \to Z \to 4\,\mu$ run I LHC searches by CMS and ATLAS  \cite{CMS:2012bw, Aad:2014wra}, Ref.~\cite{Altmannshofer:2014cfa} were able to bound $L_\mu - L_\tau$ $Z'$ with masses between $1\, \gev \lesssim M_{Z'} \lesssim 30\, \gev$ and couplings $\epsilon \gtrsim 0.04$. This bound is shown, along with the CCFR/CHARMII bound in Fig.~\ref{fig:defaultcuts}. For $M_{Z'} \gtrsim 15\,\gev$, the $pp \to Z \to 4\,\mu$ bound is more stringent.
	
	 Realizing that the LHC can place bounds on $L_\mu - L_\tau$ scenarios, in the next section, we explore how these bounds can be improved in run-II of the LHC. For starters, we will simply keep analysis the same as run-I and see how the increased energy and luminosity of run-II affect the bounds. Next, we will optimize the search by imposing some cuts to take advantage of kinematic differences between the $pp \to Z \to 4\,\mu$ events with and without a $Z'$ contribution.  As we will see, set set of cuts that optimize the search depend on the mass of the $Z'$, but in all cases the optimized searches do lead to significant improvement. Given the CCFR and run-I bounds recapped above, the target parameter region for these optimized searches is the remaining unconstrained space, roughly $\epsilon \lesssim 0.04$ (0.02) for $M_{Z'}$ heavier (lighter) than $15\, \gev$. Ideally, we would like to also cover the $L_\mu - L_\tau$ parameter space currently preferred by the $(g-2)_\mu$ anomaly,  $\epsilon \sim [0.001, 0.005]$ for $ M_{Z'} \lesssim 0.5 ~ \gev$.  
			

\section{Expected constraints from HL-LHC and future $e^+e^-$ collider}
\label{sec:analysis}


\subsection{ Extrapolating existing $pp \to Z\rightarrow 4 \mu$ searches to the HL-LHC}
\label{sec:pp_fourmu}	

In this section, we investigate  how the increase in luminosity at run-II will translate into $Z'$ reach, keeping the cuts the same as in the $8\,\tev$ analysis  \cite{CMS:2012bw, Aad:2014wra}. The cuts that CMS and ATLAS used for $pp \to Z \to 4\mu$  at LHC run-I are summarized below:

\begin{itemize}
\item [--] four isolated muons, each separated from the others by $\Delta R > 0.1$.
\item [--] the leading three leptons must satisfy $p_T > 20, 15, 8   ~  \gev$ respectively. These values are set to be as inclusive as possible while still triggerable via the dileptonic trigger. The fourth muon must satisfy the off-line id requirements: $p_T > 4\, \gev,\, |\eta| < 2.7$.  
\item [--] to select events coming from on-shell $Z$ production, the invariant mass of the sum of all four muons must satisfy $76 ~ \gev < m_{4\mu} < 106 ~ \gev$.
\item  [--] additionally, $ m_{\mu^+ \mu^-} > 4 ~ \gev$ for each pair of opposite sign muons to veto the $J/\psi$ background. 
\end{itemize}

 Through means of {\tt FeynRules}~\cite{Alloul:2013bka}, we generated a Universal Feynman rules Output (UFO) model \cite{Degrande:2011ua} for the $L_\mu - L_\tau$ Lagrangian. This model was fed into {\tt MadGraph5-aMC@NLO} \cite{Alwall:2011uj} for all simulations, including the calculation of the total width of $Z'$ for a given $\epsilon$. 

Using the cuts above and increasing the collider energy to $14\, \tev$, we can calculate $pp \to Z \to 4\mu$ rate in the SM and including the $L_\mu - L_\tau$ $Z'$. Contours of the percent difference: $100 \times (\sigma( pp \to Z \to 4\mu)_{SM + Z'} - \sigma(pp \to Z \to 4\mu)_{SM})/\sigma(pp \to Z \to 4\mu)_{SM}$ are shown below in Fig.~\ref{fig:defaultcuts} as a function of the rescaling factor $\epsilon$ and $M_{Z'}$. To get an idea how these percent differences translate into a luminosity necessary to discover a given $Z'$ scenario, we can treat $S = \mathcal L\times (\sigma( pp \to Z \to 4\mu)_{SM + Z'} - \sigma(pp \to Z \to 4\mu)_{SM})$ as the new physics ``signal" on top of the ``background" $S_0 = \mathcal L \times \sigma(pp \to Z \to 4\mu)_{SM}$ (here $\mathcal L$ is the luminosity), then define a test statistic as $S/\sqrt{S_0}$. Requiring $S/\sqrt{S_0} > 5$, we can assign a rough discovery luminosity to each of the difference contours. These luminosities are also indicated in Fig.~\ref{fig:defaultcuts}.

 \begin{figure}[h!]
 \centering
\includegraphics[width=0.6\textwidth]{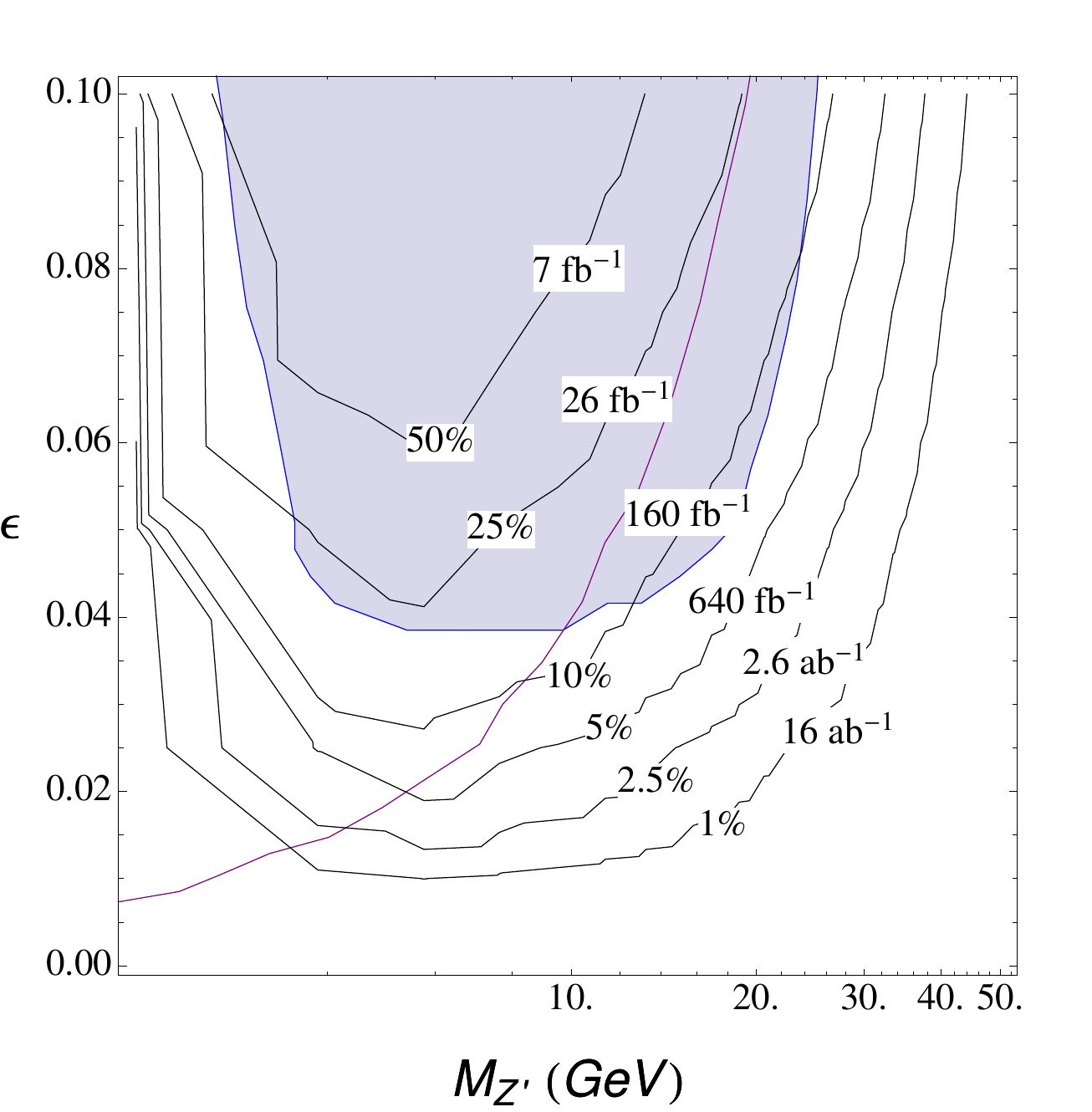} 
\caption{ The current bounds on $L_\mu -L_\tau, ~ Z'$ model is shown above. The shaded blue region represents the bounds from LHC $8~ \tev$ \cite{Altmannshofer:2014pba}, and the purple line is the CCFR bound from neutrino-trident production. The contour lines show the percent increase in the rate of cross section $100\,\times\,(\sigma(pp \rightarrow Z \rightarrow 4\mu)_{\rm Z'+ SM }/ \sigma(pp \rightarrow Z \rightarrow 4\mu)_{\rm SM} - 1)$ and the needed luminosity to get $5\,\sigma$ significance at LHC 14 TeV. The LHC bounds are derived from the conventional set of cuts for four muon production at $Z$ pole in both CMS and ATLAS  \cite{CMS:2012bw, Aad:2014wra}.}
\label{fig:defaultcuts}
\end{figure}

Repeating the $8\,\tev$ search at higher energy, we find that we can cover the $L_{\mu} - L_\tau$ parameter space down to $\epsilon \sim 0.02$ (for $M_{Z'} \lesssim 15 ~ \gev$) after the full $3\, \text{ab}^{-1}$ HL-LHC run. This is certainly an improvement, but we would like to do better. To go beyond this simple extrapolation, we need to design better searches. Specifically, we want a search that makes use of the fact that, in events with a $Z'$, two of the muons come from an on-shell, narrow particle. We present a search strategy that makes use of this feature in the next section.  As we will see, these optimized searches allow us to extend the reach of $pp \to Z \to 4\mu$ searches to smaller $\epsilon$, and will significantly decrease the luminosity needed to discover/exclude scenarios with larger $\epsilon$.

However, before exploring how to modify $pp \to Z \to 4\mu$ searches to be more sensitive to $L_\mu - L_\tau$ $Z'$, there are a few subtleties concerning our test statistic, $S/\sqrt{S_0}$. This significance estimator assumes that there is no background from continuum $pp \to 4\mu$ production (hence why we call it $S_0$ and not $B$) and ignores systematic uncertainties.  As we can see from Fig. \ref{fig:defaultcuts}, our signal tends to be only a few percents higher than SM background, while uncertainties from parton distribution functions,  soft/collinear initial state radiations (and higher order corrections in general) may be as high as 4-5\%~\cite{Watt:2012tq,Ball:2014uwa}. Thankfully, we can mitigate most of these uncertainties and non-resonant backgrounds by studying the `electron channels': $Z\rightarrow 4\,e$ and $Z \rightarrow 2e\, 2\mu$.  In the SM, the electron channels should be exactly the same as $ Z \rightarrow 4 \mu$, up to  $O(m_e/m_\mu)^2 \sim 10^{-4}$. So, by measuring these control channels we obtain a solid prediction for the SM $pp \to Z \to 4\, \mu$ rate and can therefore be sensitive to smaller deviations. Note that this is an extra advantage of studying $L_\mu - L_\tau$ model, as both electron and muon channels are affected in generic dark photon models. In practice, we will assume the systematic uncertainties on the SM $4\mu$ background are sub-percent and stick with $S/\sqrt{S_0}$ as our test statistic provided the $L_\mu - L_\tau$ effects have a signal to background ratios of a few percent or more.

\subsection{Optimized $Z'$ searches in $pp \to Z \to 4\mu$}
\label{sec:optimal}
	  As our $Z'$ is massive, one might think that the invariant masses of various opposite-sign muons $m_{\mu^+ \mu^-}$ is an effective way to distinguish signal from background.\footnote{With four muons in each event, there are four different muon pairs we can study.} For $\epsilon > 0.05$, this is true, however for smaller coupling the signal gets washed out by combinatorics. The problem is that finding the `right' muons -- the pair that reconstruct the $Z'$ mass  -- becomes challenging as $\epsilon$ gets smaller. Each of the final state events has four muons, and there are multiple combination of the leptons that can reconstruct the $Z'$ mass. We can use the kinematics of the muons to try to eliminate some of the combinatorial headache, however this depends strongly on the mass of the $Z'$. For instance, when the $M_{Z'} \ll M_{Z}$, the muons from the $Z'$ decay to tend to be soft, so we want to look at different distributions than if $M_{Z'} \sim M_Z/2$. A second issue we face when trying to pick the signal out from the background is that the signal cross section becomes quite small either as we take small $\epsilon$ or as $M_{Z'}$ approaches $M_Z/2$ (phase space suppression). Small cross sections mean that we must make a compromise between cuts that are inefficient but good discriminators and cuts with less discriminating power but that keep as much signal as possible.
	  	  
	  As the optimal analysis strategy depends on the mass of the $Z'$, we will look at a set of benchmark masses and couplings. The points we have chosen are $ M_{Z'} = 0.5,~ 10 ,~ 20, ~30,$ and $40$ GeV for $ \epsilon = 0.05, ~0.01,$ and $0.005$. As a starting point for our analysis, we require four muons satisfying $p_T > 17~ \gev, |\eta| < 2.7$ for the leading muon, $p_T > 8\,\gev, |\eta| < 2.7$ for the subleading, and $p_T > 4 ~ \gev, |\eta| < 2.7$ for the others; these cuts are motivated by the dilepton trigger thresholds and off-line muon identification requirements~\cite{CMS:2012bw, Aad:2014wra}. To insure the leptons are isolated, we demand $\Delta R_{\mu \mu} > 0.05$. Additionally, we impose the invariant mass of any pair of opposite-sign muons to be greater than twice the muon mass, $ m_{\mu \mu } \geq 0.3\,\gev$. We perform all analysis in this section at parton level for simplicity, though we have checked that our results do not change significantly if showering and hadronization are included. In the following paragraphs we give a qualitative explanation of the variables we find to be useful. The detailed cut values, along with the signal and background cut flow can be found in Appendix \ref{Appendix-A}.

	One set of variables we find useful for teasing out the signal is the transverse mass $m_T(\mu_i \mu_j)$, where
\begin{equation}
m_T^2 (\mu^+_i \mu^-_j) = 2 p_T^{\mu^+_i} p_T^{\mu^-_j} (1- \cos \Delta \phi (\mu^+_i, \mu^-_j))
\end{equation}
  and $i,j$ label the $p_T$-ranking of the lepton, separated by charge. We use a subscript $0$ for the highest $p_T$ lepton of a given charge, subscript $1$ for the second hardest, etc. One reason the  $m_T(\mu^+_i, \mu^-_j)$ are more useful variables than $m_{\mu^+_i\mu^-_j}$ is that the distributions are broader, resulting in more interference between signal and background. Depending on the $L_{\mu} - L_\tau$ parameters, the rate increase gained by using $m_T(\mu^+_i \mu^-_j)$ can overcome the fact that $m_T(\mu^+_i \mu^-_j)$ is less faithful to the actual invariant mass.  Exactly which muons enter into the transverse mass and what cuts we apply depends on $M_{Z'}$. To give the reader some idea of how the $m_T(\mu^+_i \mu^-_j)$ distributions change as we vary the $Z'$ mass and which leptons are included, the area-normalized transverse mass distributions for various $M_{Z'}$ and muon combinations are shown below in Fig.~\ref{fig:MTonshell}.  
  
For the lightest benchmark, $M_{Z'} = 0.5\, \gev$ we find that cutting on the transverse mass of the softest pair of opposite sign muons is best. The background at low $m_T(\mu^+_1, \mu^-_1)$ is dominated by $Z \to \mu^+\mu^- \gamma^* $ with the $\gamma^*$ giving the other muon pair. This background blows up when the muons from the $\gamma^*$ are soft (low $m_{\mu^+_1\mu^-_1}$ or $m_T(\mu^+_1 \mu^-_1)$) -- right where the signal lies -- so using the broader $m_T(\mu^+_1, \mu^-_1)$ distribution gives better results than $m_{\mu^+_1\mu^-_1}$. 

At $M_{Z'} = 10, 20\,\gev$, we find that cutting on a $\sim$ few $\gev$ window centered on $m_T (\mu^+_i, \mu^-_j)= M_{Z'}$ is best. However, unlike when $M_{Z'} = 0.5\, \gev$, it is less clear which leptons to include in $m_T$. For $M_{Z'} = 10\, \gev$ we find that $m_T$ of the hardest muon of one charge with the softer muon of the opposite charge and combined with the softer muons of both charges, give the best result, consistent with the picture where the $Z'$ is produced as quasi-collinear radiation from one of the initial muons. At $M_{Z'} = 20\,\gev$ and higher, mass of the $Z'$ takes up a significant amount of the available energy from the $Z$ decay, so the picture of a $Z'$ as radiation is not as useful. With less intuition on the most likely signal configurations, we must rely on Monte Carlo for guidance. For $M_{Z'} = 20\, \gev$, we find $m_T(\mu^+_1\mu^-_1)$ is best. 

For even heavier $Z'$, we can think of the signal as a nearly at-rest $Z'$ produced with back-to-back muons. In this case, the leptons from the $Z'$ decay are often the hardest leptons in the system. However, while $m_T(\mu^+_0\mu^-_0)$ does distinguish the signal from the background, we find that other cuts yield better $\sqrt{S}/S_0$. In particular, for $M_{Z'} = 30, 40\,\gev$ we find the energies of the individual leptons are more useful variables. The background is dominated by a pair of energetic leptons $E_{\mu} \sim M_Z/2$ accompanied by a pair of soft leptons originating from $\gamma^*$ radiation, while events from a heavier $Z'$ have a more even energy distribution among the leptons. This difference can be exploited either by requiring a minimum energy for the third or fourth leading lepton ($p_T$ ranked), or a maximum energy on the leading leptons. 
	
	For some masses, a cut on either the azimuthal angle between two leptons, $ \Delta \phi (\mu_i, \mu_j)$, or the separation of two leptons, $ \Delta R(\mu_i, \mu_j)$ can increase our sensitivity further. For example, for very light $Z'$ we expect the two muons from $Z'$ are very close to each other and to one of the leading leptons. As such, requiring low $ \Delta \phi (\mu_1^+, \mu_0^-)$ is a useful cut. For heavier $Z'$, requiring a minimum $\Delta R$ cut between leptons can be very useful since the bsackground is dominated by configurations with two muons that come from soft/collinear $\gamma^*$ radiation and are therefore close together. 
	
\begin{figure}[h!] 
\centering
\includegraphics[width=0.54\textwidth]{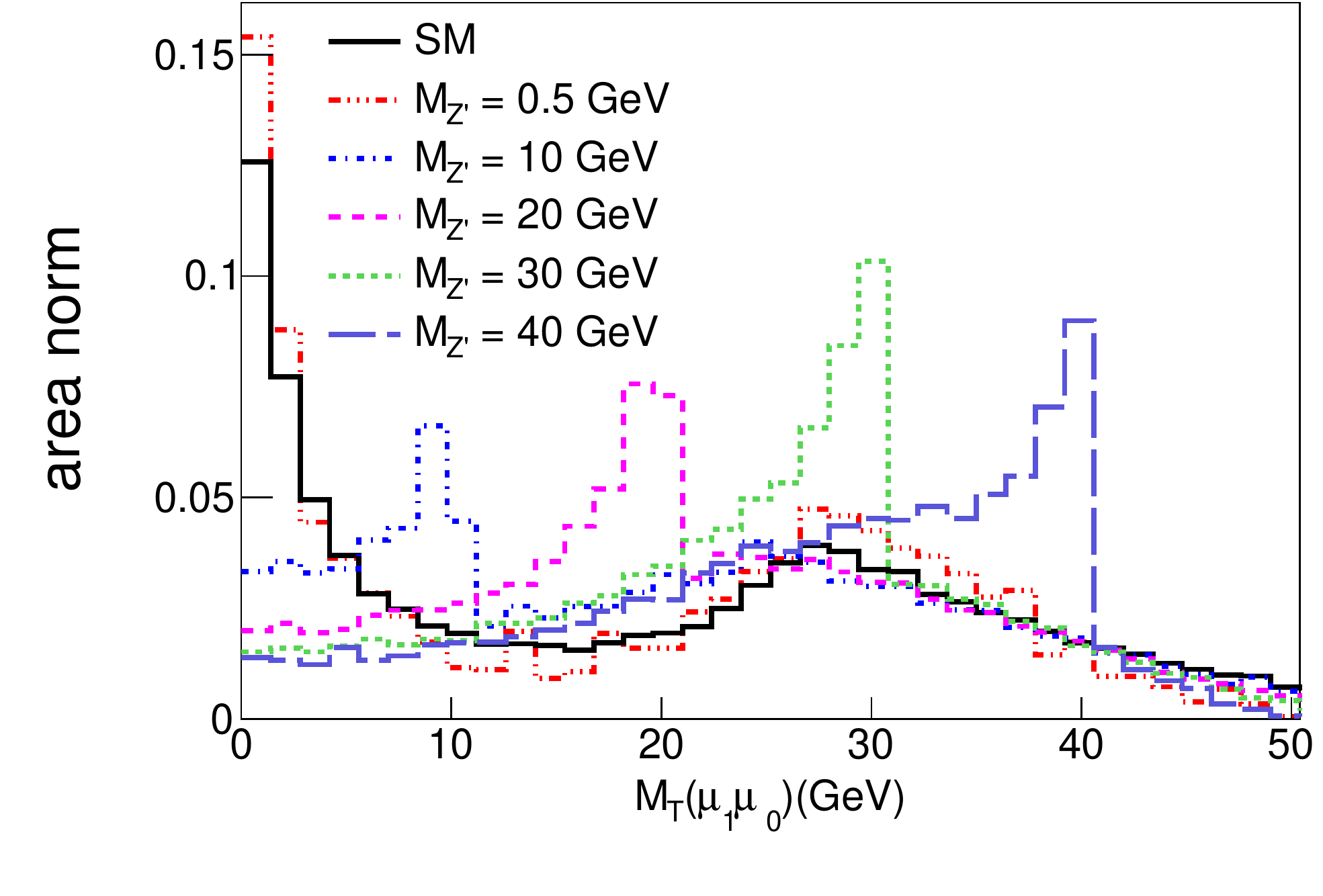}~
\includegraphics[width=0.54\textwidth]{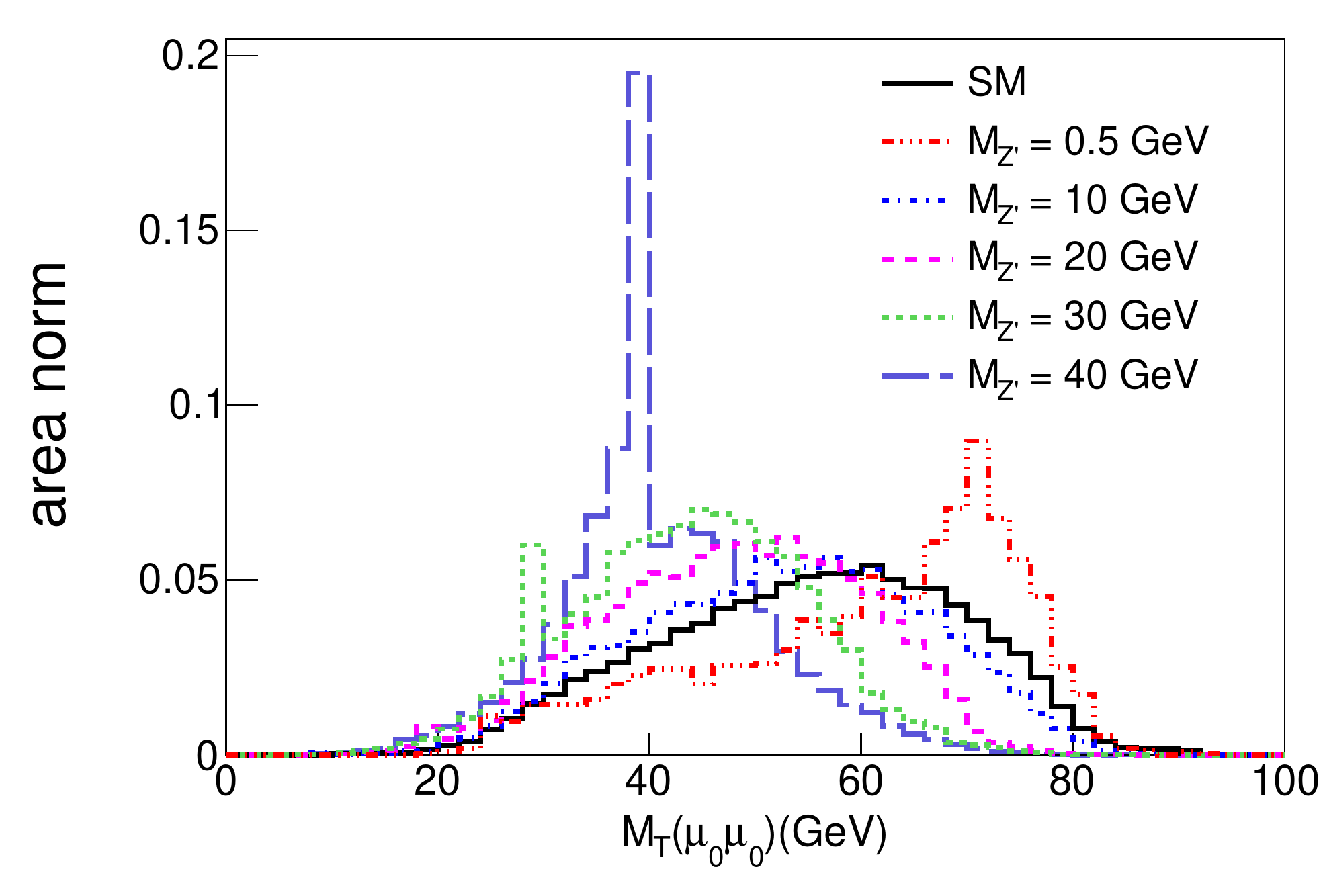}~~
 \caption{
 The distribution of the transverse mass of $\mu_1, \mu_0$ (left) and $\mu_0 ,\mu_0$ (right), for SM background and each of the $Z'$ masses. To highlight the difference in shapes, the $Z'$ plots show just the new physics contribution, i.e. $pp \to Z \to \mu^+\mu^-Z' \to 4\mu$, with no interference. All curves are area normalized. }
\label{fig:MTonshell}
\end{figure}
 
 \begin{figure}[h!]
\centering
\includegraphics[width=0.54\textwidth]{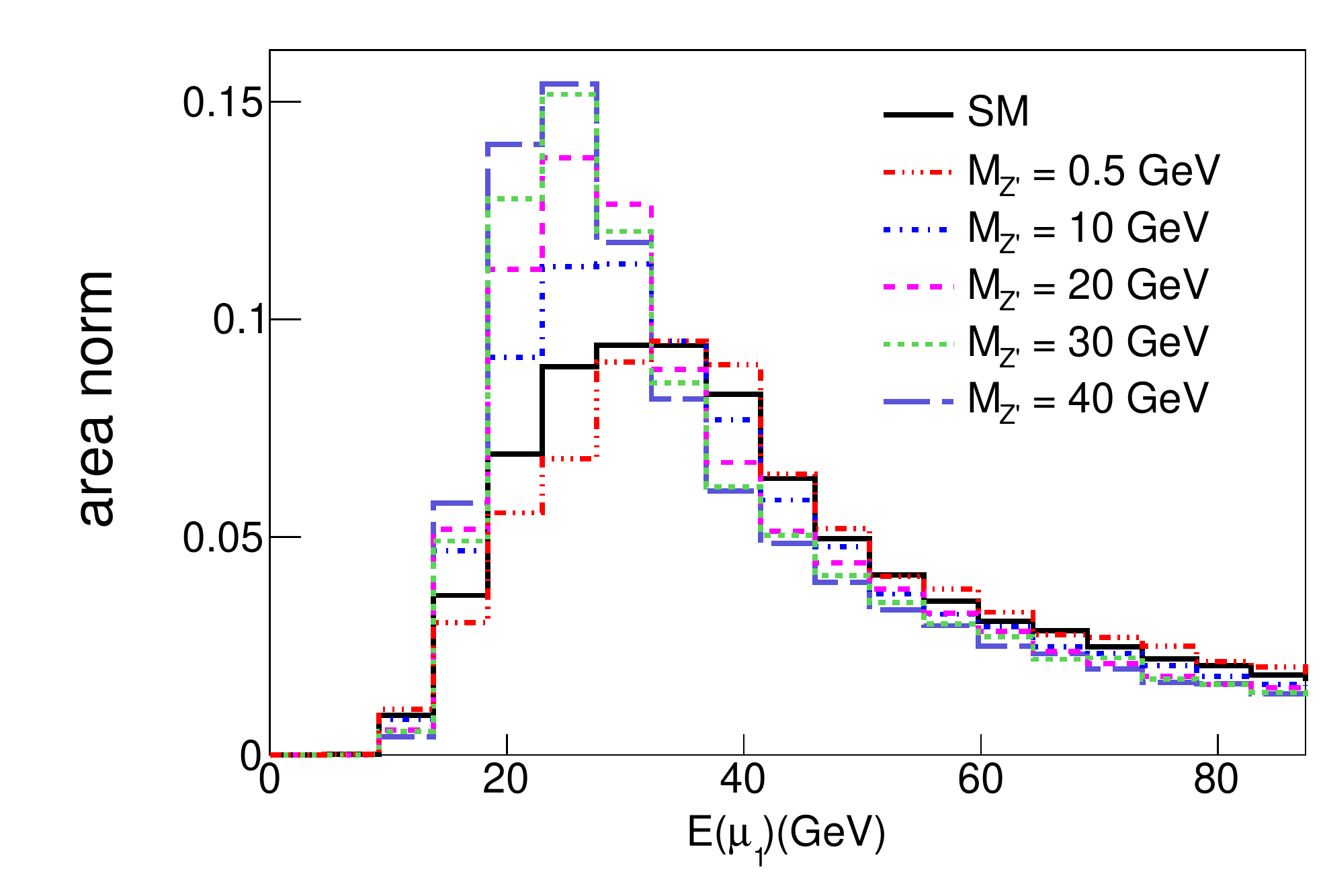}~
\includegraphics[width=0.54\textwidth]{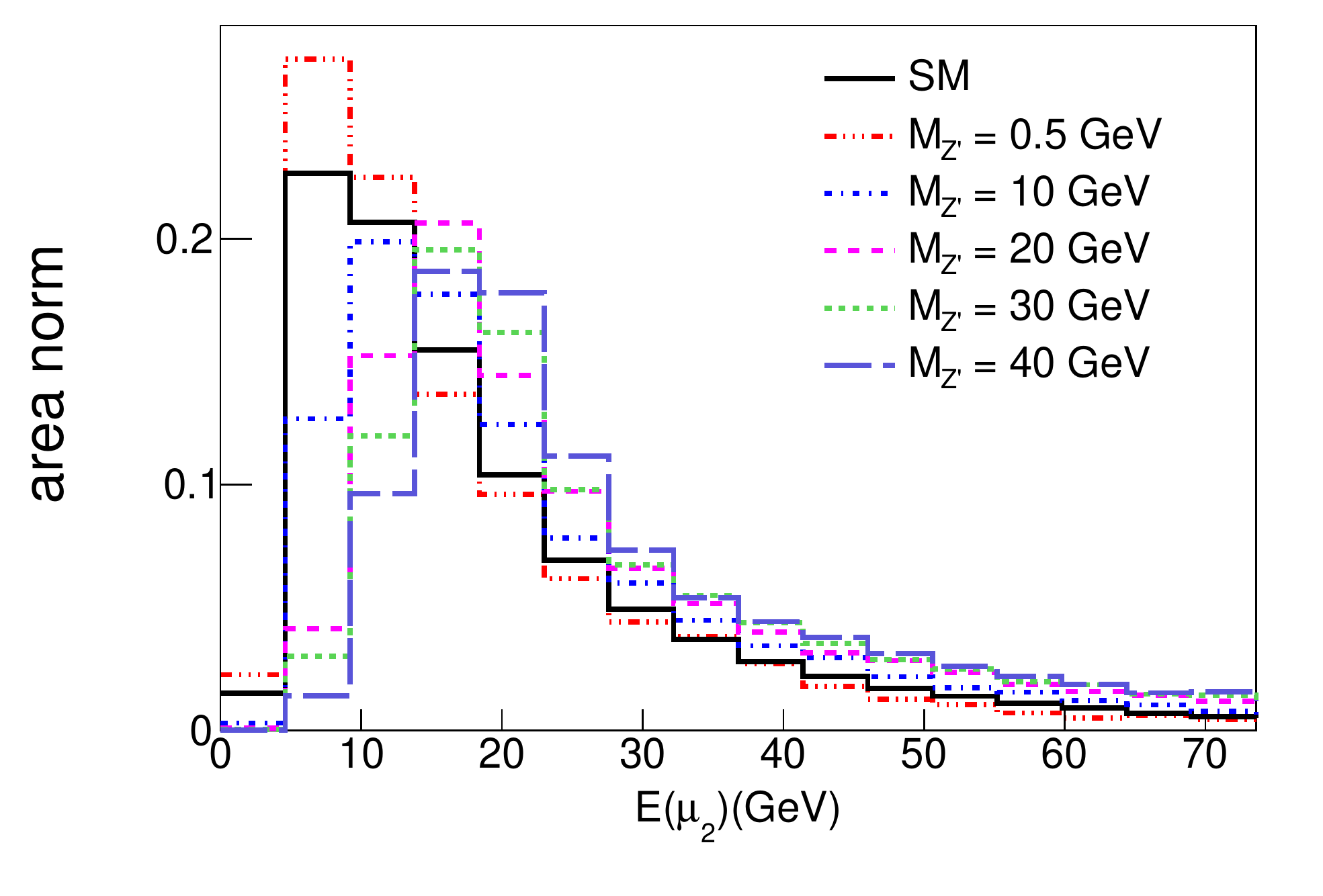}~~~
 \caption{The area-normalized distributions of the energy of second leading muon (left) and third leading muon (right), for SM background, and each of the $Z'$ masses. As in Fig.~\ref{fig:MTonshell}, the $Z'$ curves show only the resonant new physics contribution. }
\label{fig:PTonshell}
\end{figure}

 Utilizing these $M_{Z'}$-dependent cuts, with values indicated in the tables in Appendix \ref{Appendix-A}, we can compile an exclusion contour in the $\epsilon - M_{Z'}$ plane. The exclusion contour assuming the full HL-LHC luminosity of $3 ~\rm ab^{-1}$ is shown below in Fig.~\ref{fig:exclusion}, interpolating linearly between the benchmark points). With the optimized analyses presented here, we are sensitive ($4.5\,\sigma$) to $\epsilon \le 0.01$  for $M_{Z'}$ between $0.5 - 40\,\gev$. 
    
\begin{figure}[h!]
\centering
\includegraphics[width=.65\textwidth]{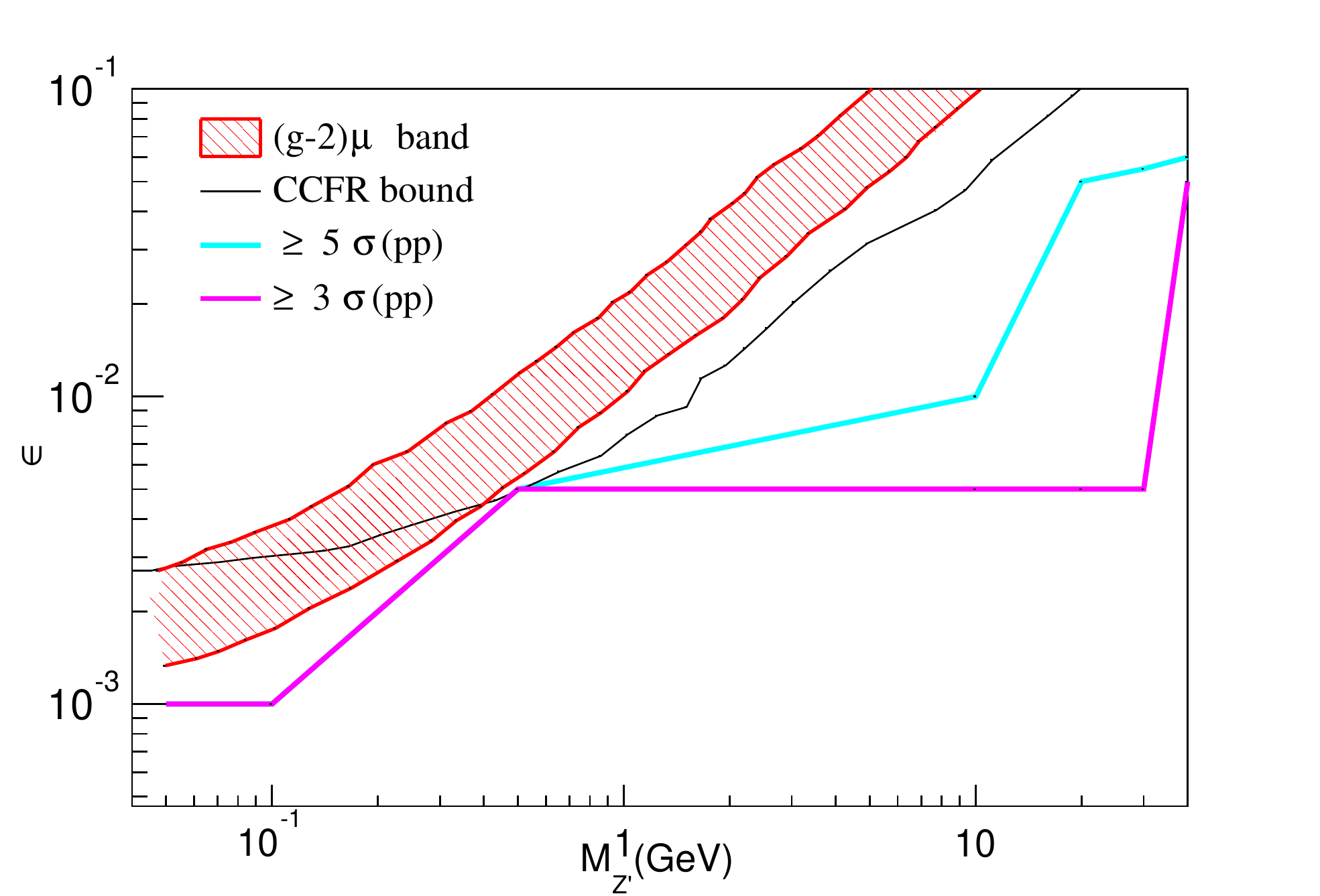}
\caption{The new bounds from our study in the LHC is shown. The blue line is the $5\, \sigma$ significance $S/\sqrt{S_0} \geq 5$, and the purple line is the $3\, \sigma$ significance $S/\sqrt{S_0} \geq 3$. The red region is the $(g-2)_\mu$ band, and the black curve shows the CCFR bounds. The HL-LHC bound is stronger than the bound from CCFR for this region in the parameter space. }
\label{fig:exclusion}
\end{figure}

\subsection{ Looking for light $Z'$ at  HL-LHC  in $ p p \rightarrow \mu^+ \mu^- ~ \slashed E_T$ }
\label{sec:pp_twomumet}			
			
	If $Z'$ is lighter than twice the muon mass it cannot be produced on-shell in the process of $pp \rightarrow 4 \mu$. 
	As the contribution from off-shell $Z'$ is suppressed by more powers of $\epsilon$, four-muon production is no longer an optimal channel for $Z'$ searches. 
	Therefore, we propose looking for $Z'$ through the process $pp \to Z \to \mu^+\mu^- Z' \to \mu^+\mu^- \nu\bar{\nu}$,  with a $\mu^+\mu^- + \slashed E_T$ final state.\footnote{This final state also captures the process where the fermions are produced in the opposite order, i.e. $pp \to Z \to \nu\bar{\nu}Z' \to \nu\bar{\nu} + \mu^+\mu^-$. However, this process only occurs through off-shell $Z'$ and so will play a negligible role.} In addition to admitting an on-shell, ultra-light $Z'$, this process has the advantage of eliminating  $pp \to Z \to \mu^+\mu^-\gamma^*$, the dominant background in the previous section. However, due to the presence of  $\slashed E_T$ in the final state, we can no longer impose the invariant mass of the leptons to be $M_Z$. Relaxing this condition introduces some new backgrounds: 	
\begin{align}
pp\rightarrow&\, \left.\tau^+ \tau^- \right|_{\text{dilepton decay}} \nonumber  \\
pp\rightarrow&\, \left.W^{+*} W^{-*} \right|_{\text{dilepton decay}} \nonumber \\
pp\rightarrow&\, Z^*\,(Z^*/\gamma^*) \nonumber \\
pp \rightarrow&\,  \mu^+\mu^-+ \text{jets} 
\label{eqn:backgrounds}
\end{align}	  
where the missing energy in the last background is assumed to come from a combination of jet mis-measurement and pileup.

 Fortunately, these backgrounds have different topology than the signal, and thus they can be removed using some careful cuts. Leptonic tau production, $p p \rightarrow \tau^+ \tau^- \rightarrow \mu^+ \mu^- ~\slashed E_T$ is by far the largest background, so we first focus on removing it.
 For this analysis, we work with fully showered and hadronized events, using {\tt PYTHIA 6.4}~\cite{Sjostrand:2006za} to decay the $\tau^{\pm}$ to $\mu^{\pm}$s and $\nu_\mu\,(\bar{\nu}_{\mu}) $.\footnote{For these searches, we set $m_{\mu} \ne 0$ but keep $m_e = 0$. } To make sure the muons are isolated, we require that the transverse energy of particles (excluding the muon) in a cone within $ \Delta R < 0.3$ of the muon to be less than $0.275$ times the $p_T$ of the muon \cite{CMS:2012bw}.
		
If the tau pair are produced from an on-shell $Z$, they will be back to back, and each $\tau$ will carry an energy of $M_Z/2$.
The $\tau^{\pm}$ are therefore boosted, and so their decay products ($\mu$ and $\nu_\mu$) tend to be collinear. Consequently,  the neutrino momenta from one tau decay partially cancels the neutrino momenta from the other tau decay, leading to relatively low $\slashed E_T$. Low $\slashed E_T$ implies that related quantities, such as the transverse mass formed from either muon with the $\slashed E_T$, will also be small.  Following this logic and exploring several variables, we find that the most efficient cut is on the transverse mass formed from the vector sum of the two leptons and the $\slashed E_T$, $m_T(\mu^+ \mu^-, \slashed E_T)$. A cut of $m_T(\mu^+ \mu^-, \slashed E_T) > 100\, \gev$ efficiently removes most of the $\tau \tau$ background without significantly affecting the signal.

The $Z \to \tau^+\tau^-$ background can be essentially eliminated after two additional cuts,  $ |\Delta \phi (\mu_0, \slashed E_T) | < 2.5 $ and $\slashed E_T > 45\, \gev$. The $|\Delta \phi|$ cut also takes advantage of the fact that the two $\tau$ in the background will be back-to-back, as the collinearity of the all objects in the event implies that the leading $p_T$ lepton must be balanced by a combination of the subleading lepton and the $\slashed E_T$.  Hence, $|\Delta \phi (\mu_0, \slashed E_T) | $ is peaked at $\pi$. The missing energy cut reduces the tau background even further, and is also useful to suppress the $p p \rightarrow \mu^+ \mu^-+X$ background. The missing energy in $p p \rightarrow \mu^+ \mu^-+X$ comes from mismeasurement and soft radiation and is therefore highly peaked at zero. We choose $\slashed E_T > 45\, \gev$ and assume this is sufficient to remove environmental backgrounds. If the high-luminosity environment proves to be so chaotic that this cut is insufficient, it could be raised without dramatically affecting our conclusions. To illustrate these cuts, the distributions for $m_T(\mu^+\mu^-, \slashed E_T), \slashed E_T$ and $\Delta \phi(\mu_0, \slashed E_T)$ are shown below in Fig.~\ref{fig:MET} for $Z \to \tau^+ \tau^-$ and SM $pp \to V^*V^* \to  \mu^+\mu^- + \slashed E_T$ processes.

\begin{figure}[h!]
\centering 
\includegraphics[width=0.48\textwidth, height=0.24\textheight]{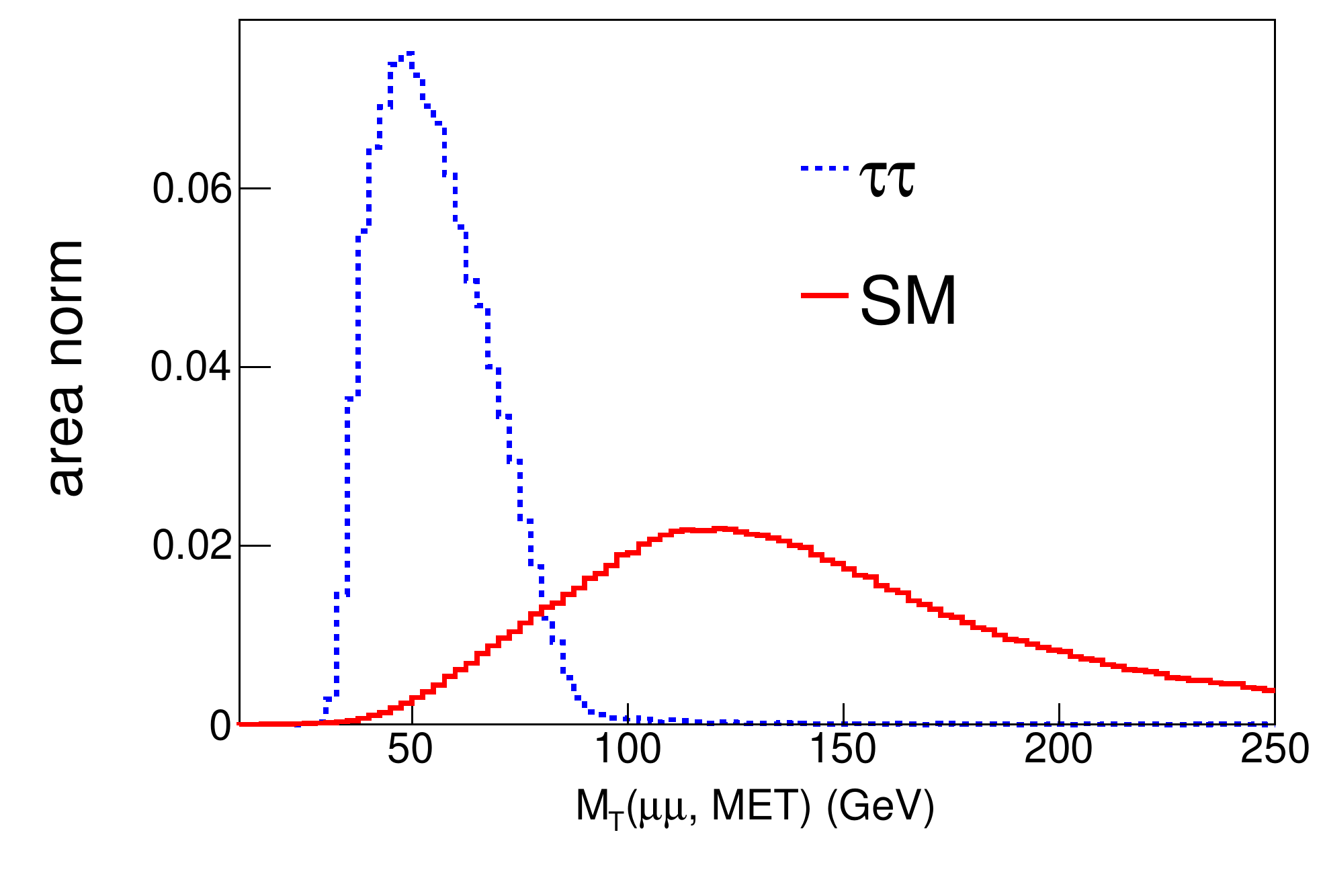} ~
\includegraphics[width=0.48\textwidth, height=0.24\textheight]{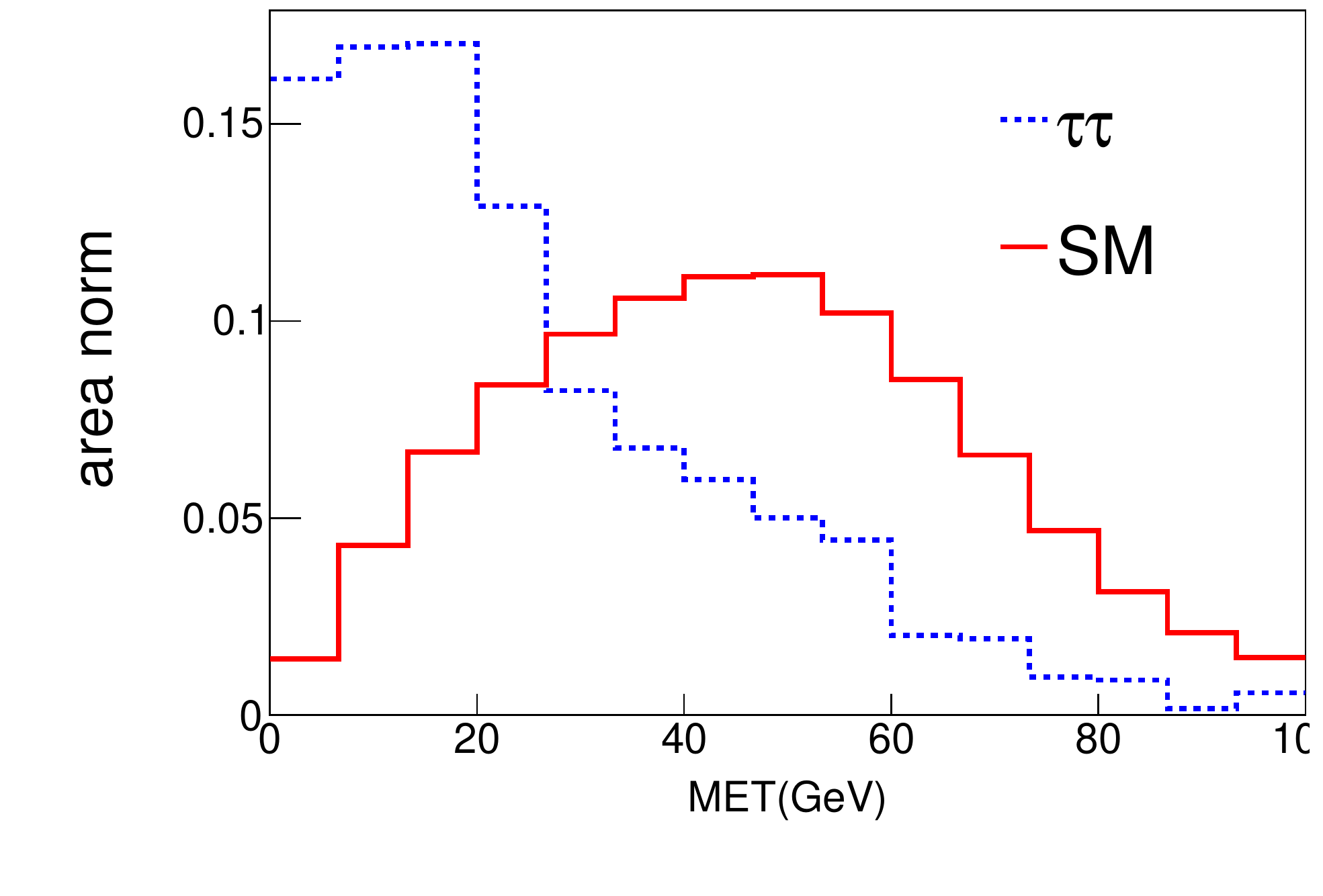} \\
\includegraphics[width=0.54\textwidth, height=0.24\textheight]{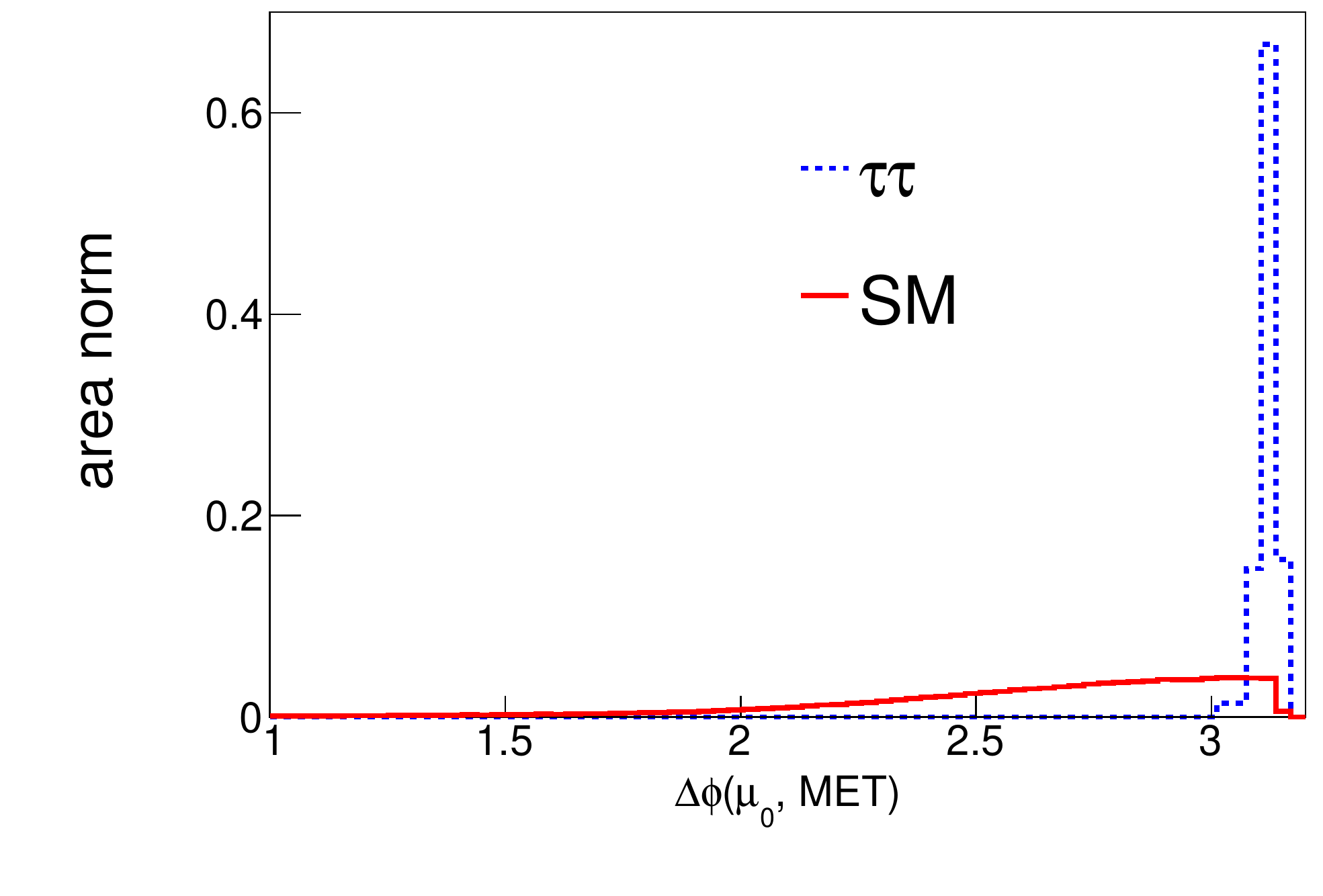} 
\caption{The series of cuts to remove $\tau \tau$ background. The blue line is the $\tau \tau$ background and SM $(pp \rightarrow \mu^+ \mu^- \nu_\mu \nu_\mu)_{\rm SM}$ production is shown with the red line. Both processes are area normalized. The first panel is the dilepton-$\slashed E_T$ transverse mass $m_T(\mu \mu, \slashed E_T)$. The second panel shows the $\slashed E_T$ after requiring $m_T(\mu \mu, \slashed E_T) > 100 ~ \gev$. Finally, the bottom panel shows the azimuthal angle between the leading lepton and $\slashed E_T$ after the $m_T(\mu\mu, \slashed E_T)$ cut and $\slashed E_T > 45\, \gev$.  }
\label{fig:MET}
\end{figure} 
	
With the $\tau \tau$ background effectively eliminated, we can focus on the residual SM backgrounds. 
The largest remaining background is  $p p \rightarrow V^* V^* \rightarrow \mu^+ \mu^- +\slashed E_T$. 
 For both $ pp \rightarrow W^{+*} W^{-*} \rightarrow \mu^+ \mu^- + \slashed E_T$ and $pp \rightarrow Z^* (Z^*/\gamma^*) \rightarrow \mu^+ \mu^- + \slashed E_T$, the invariant (and transverse) mass of the two muons is unrestricted.  Meanwhile, in the signal the two muons and neutrinos come predominantly from an on-shell $Z$, so $m_{\mu^+\mu^-}$ (and $m_{T}(\mu^+\mu^-))$ lie below $M_Z$.\footnote{A small fraction of the signal $m_{\mu^+\mu^-}$ distribution does extend beyond $M_Z$ due to $pp \to \gamma^* \to \mu^+\mu^- Z'(\nu\bar{\nu})$.} After imposing the $\slashed E_T$ cut of $\slashed E_T > 45\, \gev$, we find that the dilepton transverse mass has more discriminating power than the invariant mass (as in the $4\, \mu$ analysis). 
         	
\begin{center}		 
\begin{figure}[h!]
\begin{center}	
\includegraphics[width=0.55\textwidth]{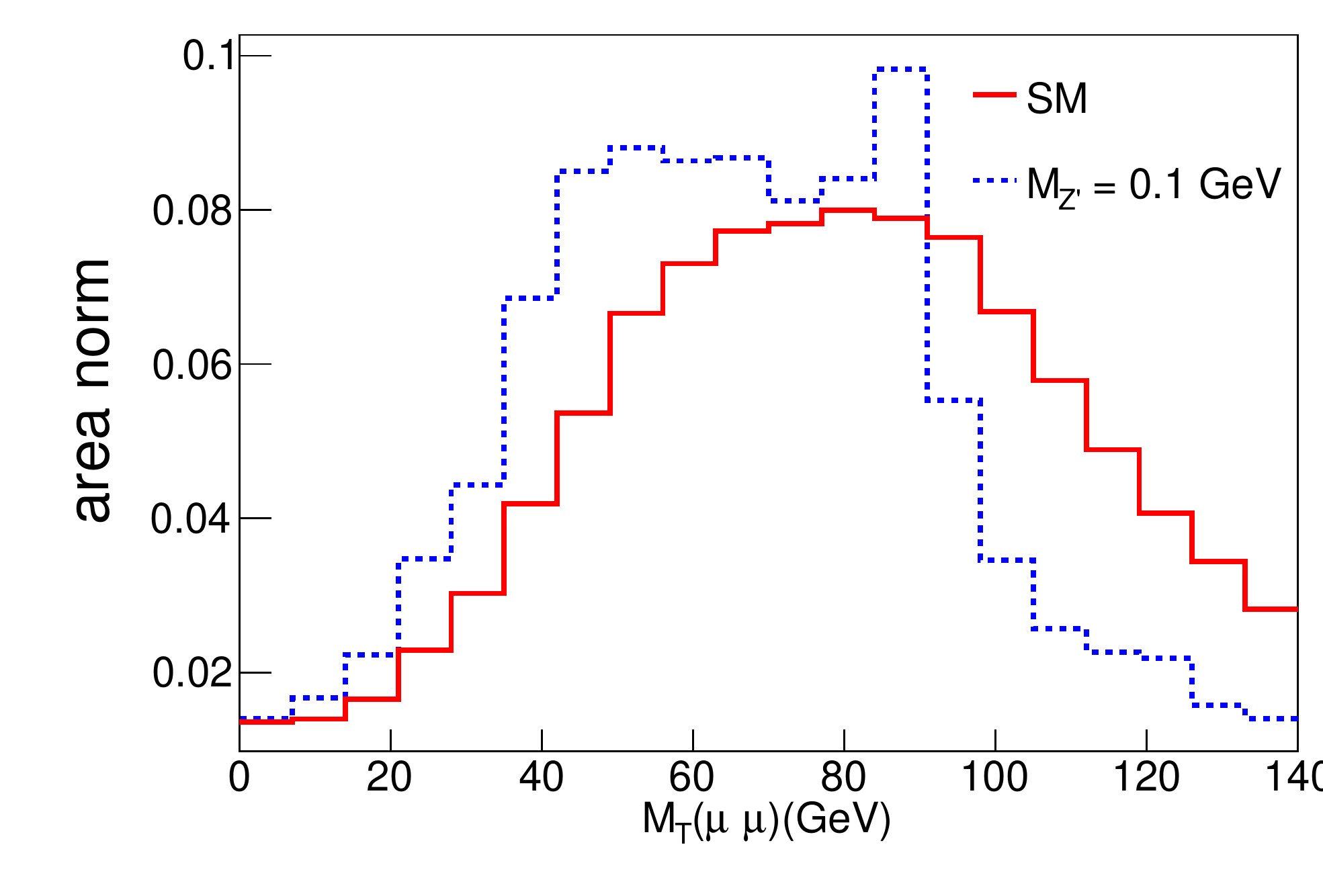} 
\end{center}
\caption {Transverse mass of the dilepton for different channels is shown here. Signal only, which is when $Z'$ is produced on-shell  ($p p \rightarrow \mu^+ \mu^- Z' \rightarrow \mu^+ \mu^- \slashed E_T$) is shown in blue for  $(M_{Z'}, \epsilon) = ~( 0.1~\gev,  0.001)$. The red line is the SM distribution $(p p \rightarrow \mu^+ \mu^- \slashed E_T)$. By requiring $m_T(\mu^+ , \mu^-) < 50 ~ \gev$, we can efficiently discriminate signal against background and get $5\sigma$ significance. }
\label{fig:MT}
\end{figure} 
\end{center}

We considered two benchmark signal points: $M_{Z'} = 0.1~\gev, \epsilon = 0.001$, and $M_{Z'} = 0.05~\gev, \epsilon = 0.001$. Imposing the $m_T(\mu^+\mu^-, \slashed E_T), \Delta \phi(\mu_0, \slashed E_T)$ and $\slashed E_T$ cuts mentioned above, along with a cut on the maximum transverse mass of the two muons $m_T(\mu^+,\mu^-) < 50\, \gev$, we find that a 5$\,\sigma$ significance\footnote{As in the $4\mu$ channel, we find the transverse mass of muons pairs gives better results than the invariant mass} can be achieved with an integrated luminosity of  $2.6~ \rm ab^{-1}$. We find that further cuts can increase the signal-to-background ratio, but necessitate an increase in discovery luminosity. These $pp \to 2\mu+ \slashed E_T$ bounds are included in Fig.~\ref{fig:exclusion} and appear in the lower left region of the plot. The details of the cuts and the cut flow for the different signal points can be found in Appendix \ref{Appendix-A}. Combining the $pp \to \mu^+\mu^- + \slashed E_T$ channel with the results from $pp \to 4\, \mu$, we can explore the region where $L_\mu - L_\tau$ models can explain $(g-2)_{\mu}$.

\subsection{ Optimized Searches in $ e^+ e^- \rightarrow Z \rightarrow 4 \mu$, with $ E =92 ~ \rm GeV$}
\label{sec:ee_fourmu}

While run-II of the LHC has just begun, it is nevertheless worthwhile to look ahead to the prospects for future colliders. In particular, it is interesting to study how the LHC fares in 
$L_\mu - L_\tau$ $Z'$ discovery when compared with a next generation $e^+e^-$ $Z$-factory. Proposals for future $Z$-factories include TLEP, a circular lepton collider which can reach energies $90~ \gev - 350~ \gev$ and beyond  \cite{Koratzinos:2013ncw}, and CEPC, a circular lepton collider based in China with energy up to $240 ~ \gev$ and a projected luminosity of $2.6~ \rm ab^{-1}$ \cite{Zimmermann:2014qxa}.

As the $L_\mu - L_\tau$ $Z'$ does not couple to electrons, $Z'$ are produced at an $e^+e^-$ collider in exactly the same fashion as the LHC, i.e. $e^+e^- \to Z/\gamma^* \to \mu^+\mu^- Z'$. While the production mechanism in the two setups is similar, the cut strategy changes as we shift from the LHC to a lepton collider. One reason for the shift is that the collisions in an $e^+e^-$ collider are much cleaner and the center of mass energy is precisely known. For our studies we assume $\sqrt s = 92\, \gev$, with negligible uncertainty from bremsstrahlung.\footnote{For TLEP, the systematic uncertainty is expected to be as low as 100 KeV at the $Z$ pole~\cite{Fan:2014vta}.} The second reason the cuts change is that the interference between the SM and $Z'$ production is sensitive to the $Z/\gamma$ charge of the initial particle.\footnote{Lepton colliders have the possibility of polarized beams, but we ignore this possibility here.} 
		 
Focusing on the $4\,\mu$ final state, the optimal cuts will depend sensitively on $M_{Z'}$, so we will study several signal benchmarks. The cleanliness of the $e^+e^-$ environment allows us to explore a larger region of $Z'$ parameter space. We will use the same signal masses as before $M_{Z'} = 0.5, ~ 10, ~ 20, ~ 30, ~40\,\gev$ but consider smaller coupling, $\epsilon = 0.01, 0.005, 0.001$.  Since there are no established triggers for a future lepton collider, we use same the dilepton trigger and base cuts listed in Sec.~\ref{sec:optimal} and, as in our hadron collider analysis, we will judge the effectiveness of a set of cuts by $S/\sqrt{S_0}$, where  $$S = \mathcal L\times (\sigma( e^+ e^- \to Z \to 4\mu)_{SM + Z'} - \sigma(e^+ e^- \to Z \to 4\mu)_{SM})$$ and $$S_0 = \mathcal{L} \times(\sigma(e^+ e^- \to Z \to 4\mu)_{SM}).$$ The integrated luminosity we will use in this study is $2.6\, \text{ab}^{-1}$, the conservative estimate of the total TLEP dataset~\cite{ Ruan:2014xxa, Zimmermann:2014qxa}.\footnote{ The expected TLEP luminosity for 240 GeV is $10 ~ \rm ab^{-1}$ and for 350 GeV is $2.6~ \rm ab^{-1}$. }

	For lighter $M_{Z'}$, multiple combinations of the leptons can reconstruct the $Z'$ mass, causing a combinatorial effect. This combinatorial background made the di-muon invariant masses ineffective variables for small $\epsilon$ at the LHC. However, at an $e^+e^-$ Z-factory, the effect from including a $Z'$ (for fixed $\epsilon$ and $M_{Z'}$) is both larger and cleaner than at a hadron collider -- increasing the range of utility of $m_{\mu^+\mu^-}$. The $Z'$ effect is larger (compared to a $q\bar q$ initiated event) because the leptons are uncolored and because the $Z-\ell-\ell$ coupling is slightly larger than the $Z-q-q$ coupling, and it's cleaner because the center of mass energy is precisely known and does not require convolution with parton distribution functions. 
	
	Cutting on the invariant mass of the leading negative muon plus the subleading positive muon (or charge conjugate), $m_{\mu_1^+ \mu_0^-}$, we find $S/\sqrt{S_0} > 5$ for $M_{Z'} = 0.5 ~ \gev$, $ \epsilon = 0.001$ after $2.5 ~ \rm ab^{-1}$ of luminosity. This particular combination of leading and subleading muons is useful since light $Z'$ will be emitted very close to one of the leading two muons, and the subleading muon in the signal tends to have larger energy than the background. 	
	 
	 Moving to larger $Z'$ masses, for $M_{Z'} = 10 ~ \gev$ and $20$ GeV, we find the most efficient variable to cut on is the transverse mass of two muons $m_T(\mu^+_i,\mu^-_j)$, with optimal cut values close to $M_{Z'}$. The combination of the muons that work for each of these masses are different and are shown in Appendix \ref{Appendix-C}. With the $m_T(\mu^+_i,\mu^-_j)$ cut alone, we find a 4$\sigma$ significance for $\epsilon = 0.001$ may be achieved after $2.6\, \text{ab}^{-1}$.  Further cuts can increase $S/S_0$ but reduce the signal cross section so much that $S/\sqrt{S_0}$ (at $2.6\, \text{ab}^{-1}$) suffers. 
	 
	 For heavier $Z'$ masses, trying to capture the two muons from the $Z'$ decay with an invariant mass or transverse mass cut results in too low of a cross section, so we have to explore other combinations.  For $M_{Z'} = 30~ \gev$, we find that requiring $m_T(\mu_0^+ \mu_0^-) \sim  M_{Z} - M_{Z'}$ is the most efficient cut. This cut, combined with a cut on the separation between two of the leptons, gets us $3.5\,\sigma$ significance with $2.6~ \rm ab^{-1}$ integrated luminosity. For the highest mass we consider, $M_{Z'} = 40$ GeV, requiring $m_T(\mu_0^+ \mu_0^-)$ to be around  $ M_{Z} - M_{Z'}$ is still the best variable, however the signal cross section is so low that we cannot get $S/\sqrt S_0 = 3$ after $2.6\, \rm ab^{-1}$. 
		
\begin{figure}[h!]
\centering
\includegraphics[width=.65\textwidth]{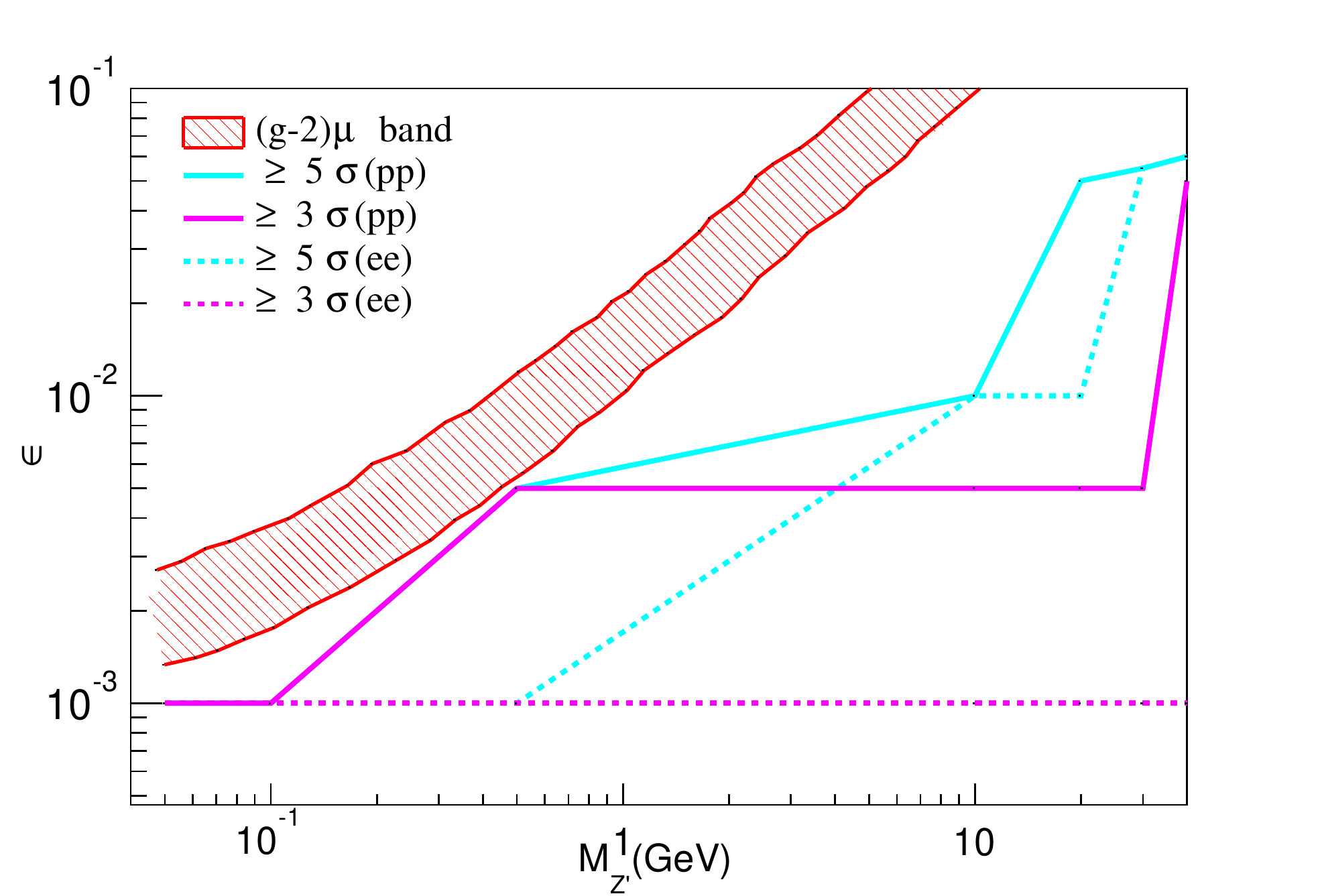}
\caption{Bounds in the $\epsilon, M_{Z'}$ plane on the $L_\mu - L_\tau$ model from the HL-LHC and future $e^+e^-$ $Z$ factory. The blue (purple) dashed line corresponds to the lepton collider $S/\sqrt{S_0} \geq 5$ $(S/\sqrt{S_0} \geq 3)$ region, and the solid blue (purple) lines are the bounds we derived from the hadron collider for $\sigma \geq 5 ~ (\sigma \geq 3)$. The red region is the $(g-2)_\mu$ band. For the luminosities we have assumed ($3\, \rm ab^{-1}$ for $pp$, $2.6\, \rm ab^{-1}$ for $e^+e^-)$, the limits from the $Z$ factory are stronger for $0.5\, \gev \le M_{Z'} \le M_Z/2$.}
\label{fig:exclusionee}
\end{figure}

	Our $e^+e^-$ $Z'$ search projections are compiled in Fig.~\ref{fig:exclusionee} below and compared to our projections from the HL-LHC. Assuming $2.6\,\rm ab^{-1}$ of luminosity, we are able to probe couplings at the $\epsilon = 0.001$ level at $3\, \sigma$ for the entire mass range of interest. Notice that the $e^+e^-$ searches in Fig.~\ref{fig:exclusionee} only cover $M_{Z'} > 0.5\, \gev$. \\
	
For $M_{Z'} < 2m_\mu$, we have shown that HL-LHC can already explore the parameter space to $\epsilon = 0.001$. However, it is still interesting to study the capabilities of a future lepton collider in this region of parameter space. As discussed in Sect.~\ref{sec:pp_twomumet}, for this range of $Z'$ masses it is better to look for $Z' \to \nu\bar{\nu}$, i.e. in  $e^+ e^-\rightarrow \mu^+ \mu^- \slashed E_T$.

With only two leptons in the event, there are fewer kinematic handles and we no longer have the luxury of ignoring all backgrounds except SM multi-lepton production. However, one tool we do have at our disposal is the recoil mass technique~\cite{ Bai:2000pr}. Specifically, assuming the $Z'$ ($\to \nu\bar{\nu})$ are always created on-shell, we can derive $Z'$ mass as a function of measurable or known parameters: 
		\begin{equation}
		M_{Z'}^2 \equiv M_{V}^2(\mu^+ , \mu^-) =   s + m_{\mu^+ \mu^-}^2 - 2 \sqrt{s} E_{\mu^+ \mu^-},
		\end{equation}
		 where $s= E_{\rm cm}^2$, and $m_{\mu^+ \mu^-}$  and $ E_{\mu^+ \mu^-}$, correspond to the mass and energy of the two muons that are not from $Z'$. Applied to our scenario, the recoil mass will exhibit a sharp peak for the signal, even though the $Z'$ decay invisibly. Meanwhile, backgrounds such as $e^+e^- \to Z/\gamma^* \to \tau^+\tau^- \to \mu^+\mu^- + \slashed E_T$ lead to broad and featureless recoil mass distributions. This difference is illustrated below in Fig.~\ref{fig:recoilmass}. Unfortunately, despite this nice kinematic discriminant, we find that we cannot achieve $S/\sqrt{S_0} \geq 5$ given $2.6\, \rm ab^{-1}$. The biggest hurdle here is that the $ \mu^+ \mu^- \nu_\ell \bar \nu_\ell$ production at the lepton collider is much smaller than the LHC, such that any cut which removes the $\tau \tau$ background degrades the signal too much.
		 	 
 \begin{figure}[h!]
\centering
\includegraphics[width=.55\textwidth]{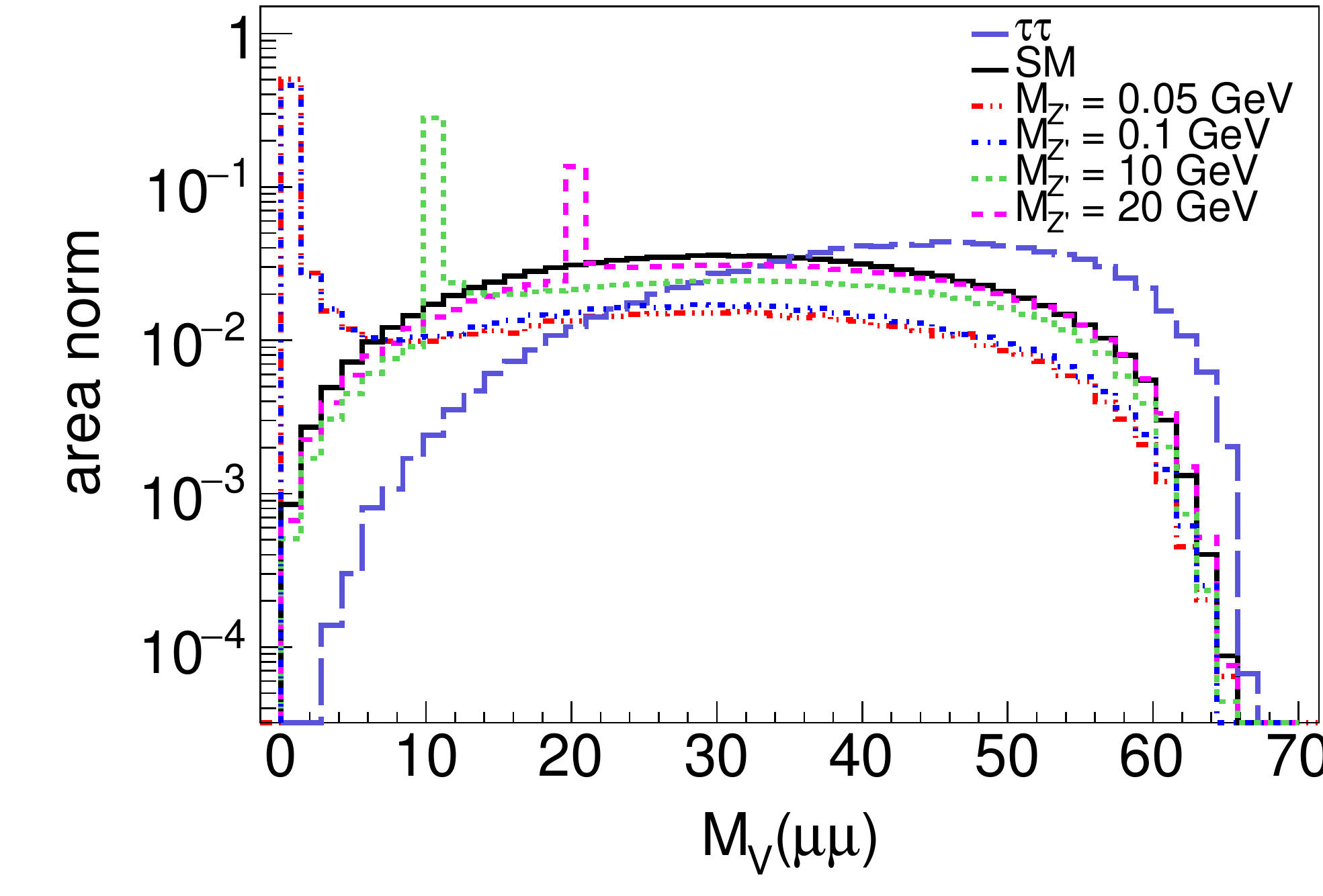}
\caption{The recoil mass for signal (resonant new physics piece only) with various $Z'$ mass and $\epsilon = 0.005$ are shown along with the recoil mass of the $\tau \tau$ background and other SM backgrounds. As we can see, there is a very distinct peak at $M_{Z'}$ in the signal. If the cross sections of the signal were not too low, a cut on recoil mass would have efficiently separated the light $Z'$ signal from both the $\tau \tau$ and $e^+e^- \to V^*V^* \to 2\,\mu + \slashed E_T $ backgrounds.}
\label{fig:recoilmass}
\end{figure}
			
\section{Discussion}
\label{sec:discuss}

In this work we have presented some strategies to find $L_\mu - L_\tau$ $Z'$ at the HL-LHC and future lepton colliders. The $L_\mu - L_\tau$ model is a simple extension of the SM, and one that is particularly difficult to bound since the new physics does not couple to electrons or quarks at tree level. Additionally, for certain values of the coupling and $Z'$ mass, $L_\mu - L_\tau$ models have been proposed as a possible way to explain the longstanding anomaly in $(g-2)_{\mu}$.

For $2\, m_{\mu} < M_{Z'} < M_Z/2$, we showed how the rare process $pp \to Z \to 4\,\mu$ can be used to set bounds on the $Z'$. The four muon final state is well understood, and one can use the related channels $pp \to Z \to 4e, 2e\,2\mu$ as background control samples to reduce systematics. We assume the systematic uncertainties can be reduced via control sample measurements to the sub-percent level, so any $Z'$ effect with signal to background ratio of a few percent or more is considered potentially visible. Using the invariant masses $m_{\mu^+_i\mu^-_j}$ or transverse masses $m_T(\mu^+_i, \mu^-_j)$ of opposite sign muon pairs as discriminating variables, we find the HL-LHC can exclude -- at $3\, \sigma$ --  the parameter space $M_{Z'} > 0.5\, \gev$ and $\epsilon > 0.005$ (in a convention where the coupling of $Z'$ to muons is $g_1\, \epsilon$) after $3\, \rm ab^{-1}$. Extending this study to a future $e^+e^-$ Z-factory, we can exclude $\epsilon > 0.001$  for the same range of $Z'$ masses.

To study $Z'$ lighter than $2\,m_{\mu}$, we proposed a search in the $2\,\mu + \slashed E_T$ final state. Even though several new and potentially large backgrounds appear when we consider this final state, we find these can be safely removed using a combination of the missing energy, the transverse mass of the dilepton plus $\slashed E_T$ system ($m_T(\mu^+\mu^-, \slashed E_T)$), and the azimuthal angle between the leading muon and the $\slashed E_T$. Incorporating the $2\,\mu + \slashed E_T$ channel, we can extend the $3\, \sigma$ exclusion limit to $M_{Z'} < 2\, m_{\mu},\, \epsilon > 0.001$, which completely covers the parameter region where the $L_\mu - L_\tau$ model can explain the $(g-2)_{\mu}$ discrepancy. For these ultra-light $Z'$, the cross section (after cuts to remove background) at a lepton collider is too small to be useful.

 \subsection*{Acknowledgments}

 We would like to thank Joe Bramante, Nirmal Raj, Bryan Ostdiek and Ciaran Williams for their insightful comments and suggestions.  The work of AM was partially supported by the National Science
Foundation under Grant No.~PHY-1417118.  This research was supported in part by
the Notre Dame Center for Research Computing through computing resources.

\section{Appendix-A}
\label{Appendix-A}

This section shows the cut flow for the search $pp \rightarrow Z \rightarrow 4 \mu$. The base cuts as defined in Section \ref{sec:analysis} are the following:

\begin{itemize}
\item [--] four isolated muons, each separated from the others by $\Delta R > 0.05$.
\item [--] di-lepton trigger: the leading two leptons must satisfy $p_T > 17, 8  ~  \gev$ respectively. All muons also have to satisfy $p_T(\mu_i) > 4 ~ \gev$ and $ \eta(\mu_i) < 2.7$.
\item [--]  on-shell $Z$ production: the invariant mass of the sum of all four muons must satisfy $76 ~ \gev < m_{4\mu} < 106 ~ \gev$.\end{itemize}

The benchmark points for this analysis are $M_{Z'} = 0.5 , 10, 20 , 30 , \rm and ~ 40 ~ \gev$, for the $ \epsilon = 0.05,~ 0.01,$ and $0.005$.  In tables~\ref{tab:ptfive}-\ref{tab:40gev} below, the values in the square brackets are $100 \times (\sigma(pp \rightarrow 4\mu)_{\rm SM+Z'} - \sigma(pp \rightarrow 4\mu)_{\rm SM} )/   \sigma(pp \rightarrow 4\mu)_{\rm SM}$. The significance after $3 ~ \rm ab^{-1}$ is shown in the last row.  Throughout all appendices, we will use $S_0$ to indicate the SM rate for the process of interest ($pp \to Z \to 4\mu, \to 2\mu + \slashed E_T$, etc.).

\subsubsection{ $pp \to Z \to 4\mu,\, M_{Z'} = 0.5~\gev$}

For an on-shell light $Z'$, with mass as light as $M_{Z'} = 0.5~\gev$, the two muons coming from $Z'$ are expected to be the subleading positive and negative muons. Moreover, from the topology of the signal, we expect that the subleading muons to be close to one of the (positive or negative) leading muons. As shown in table \ref{tab:ptfive}, with a combination of $ m_T(\mu_1^+ \mu_1^-)$ and  $ \Delta \phi (\mu_1^+ \mu_0^-) $, we can achieve the desired significance. 

\begin{centering}
\begin{table}[H]
\makebox[\textwidth]{
\begin{tabular}{c c  c c c c c  | c  c  }
\hline
\hline
\multicolumn{8}{c}{Cross section [fb]}  \\
cut & & $\epsilon = 0.05 $ & & $\epsilon  = 0.01 $ && $ \epsilon  =0.005$&& SM \\
0. Basic cuts && 43.3 [$(22.1 \pm 0.7)$ \%] && 35.9 && 35.7 && $35.6 \pm 0.04 $ \\
1. $ m_T(\mu_1^+ \mu_1^-) < 0.5~ \gev$ && 15.0 [$(82\pm 1.0)$\%] && 8.5 [$(3.5\pm  0.2)$\%] && 8.4[$(2.0 \pm 0.2)$\%] &&$8.2 \pm 0.02$\\
2. $ \Delta \phi (\mu_1^+ \mu_0^-) < 0.15$ && 1.94 [$(204\pm 3)$\%] && 1.01 [$(11.0 \pm  0.6)$\%] && 1.00 [$(10.0 \pm 0.6)$\%] &&$0.91 \pm 0.01$\\
\hline
$\left. (S - S_0)/\sqrt{S_0}\right|_{\mathcal{L} = 3\, \rm ab^{-1}} $ &&63.0 && 11.0 &&5.2&&\\
&& && && \\
\hline
\hline
\end{tabular}}
\caption{The cross section after each cut and its uncertainty for $M_{Z'}  = 0.5 ~ \gev$ in $pp$ collider. }
\label{tab:ptfive}
\end{table} 

\end{centering}



\subsubsection{$pp \to Z \to 4\mu,\, M_{Z'} = 10~ \gev$}

In this case also, multiple combination of $ \mu_i \mu_j$ can reconstruct $M_{Z'}$. Therefore, we require $m_T(\mu_i \mu_j) \sim M_{Z'}$, where $ \mu_i $ and $ \mu_j$ are any leptons except for the combination $ m_T (\mu_0^+ \mu_0^-)$. An extra cut on the separation between two leptons can both reduce the photon background and favor the topology of on-shell $Z'$ (see table \ref{tab:10gev}).


\begin{centering}
\begin{table}[H]
\makebox[\textwidth]{
\begin{tabular}{c c  c c c c c | c  c  }
\hline
\hline
\multicolumn{8}{c}{Cross section [fb]}  \\
cut & & $\epsilon = 0.05 $ & & $\epsilon  = 0.01 $ && $ \epsilon  =0.005$  && SM \\
0. Basic cuts &&39.0 && 35.8&& 35.7&& $35.6 \pm 04 $ \\
1. $ \left\{ \begin{matrix} ~~~~  6 ~ \gev <m_T(\mu_1^+ \mu_1^-) < 14~ \gev\\ \rm or~  6 ~ \gev< m_T(\mu_1^+ \mu_0^-) < 14~ \gev \\  \rm or~  6 ~ \gev <m_T(\mu_0^+ \mu_1^-) < 14~ \gev \end{matrix} \right.$
 &&13.9 [$( 18.3 \pm 0.2)$\%] && 12.0  [$(2.0 \pm  0.2 )$\%] && 11.78 [$( 0.2 \pm 0.2 )$\%] &&$11.76 \pm 0.02$\\
 2. $1 < \Delta R (\mu_0 \mu_3) < 2.5  $ && 3.20 [$( 35.2 \pm 0.3)$\%] && 2.52 [$(6.4 \pm  0.3 )$\%] && 2.46[$( 3.8 \pm 0.3)$\%] &&$2.37\pm 0.01$\\
\hline
$\left. (S - S_0)/\sqrt{S_0}\right|_{\mathcal{L} = 3\,\rm ab^{-1}} $ && 29.5 && 5.3 && 3.2 &&\\
&& && &&\\
\hline
\hline
\end{tabular}}
\caption{The cross section after each cut and its uncertainty for $M_{Z'}  = 10 ~ \gev$ in $pp$ collider. }
\label{tab:10gev}
\end{table}
\end{centering}



\subsubsection{$pp \to Z \to 4\mu,\, M_{Z'} = 20~\gev$}

For this case, the cut on $ m_T(\mu_1^+ \mu_1^-)$ is the most optimal cut. After this cut, as shown in table \ref{tab:20gev}, the cross section becomes very low and any extra cut will require higher luminosity than $3 ~ \rm ab^{-1}$ to get the same significance. 

 \begin{centering}
\begin{table}[h!]
\makebox[\textwidth]{
\begin{tabular}{c c  c c c c c  | c  c  }
\hline
\hline
\multicolumn{8}{c}{Cross section [fb]}  \\
cut & & $\epsilon = 0.05 $ [fb]  & & $\epsilon  = 0.01 $ [fb]  && $ \epsilon  =0.005$ [fb]    && SM [fb]  \\
0. Basic cuts && 37.3 && 36.1 && 35.6   && $35.6\pm 0.04$ \\
1. $17 ~ \gev < m_T(\mu_1^+ \mu_1^-) < 20~ \gev$ && 1.38 [($89.1 \pm 0.6 $)\%] &&0.76 [($8.6  \pm 0.6$ )\%] && 0.75[($7.0   \pm 0.6$ )\%]&&$0.70 \pm 0.005 $\\
\hline
$\left. (S - S_0)/\sqrt{S_0}\right|_{\mathcal{L} = 3\, \rm ab^{-1}} $ &&41.2 && 4.0&&3.2  &&\\
&& && &&\\
\hline
\hline
\end{tabular}}
\caption{The cross section after each cut and its uncertainty for $M_{Z'}  = 20 ~ \gev$ in $pp$ collider. }
\label{tab:20gev}
\end{table}
\end{centering}


\subsubsection{$pp \to Z \to 4\mu,\, M_{Z'} = 30\, \gev$}

The leptons coming from on-shell $M_{Z'} = 30\, \gev$ have energies around 15 GeV. A cut on the energy of the third-hardest muon, $E^{\mu_2}$ combined with a cut on the separation between two leptons (table \ref{tab:30gev}), we increase our sensitivity to spot $Z'$ for this mass.


 \begin{centering}
\begin{table}[H]
\makebox[\textwidth]{
\begin{tabular}{c c  c c  c c  | c  c  }
\hline
\hline
\multicolumn{8}{c}{Cross section [fb]}  \\
cut  & $\epsilon = 0.05 $ [fb]  && $\epsilon  = 0.01 $ [fb]  && $ \epsilon  =0.005$ [fb]  && SM [fb]  \\
0. Basic cuts & 36.2&& 35.6 && 35.6 &&$35.5 \pm 0.1$ \\
1. $ 9~ \gev < E^{\mu_2} < 15~ \gev$ & 9.63   [($ 2.0 \pm 0.2$)\%] &&9.47 [($0.3  \pm 0.2$ )\%] && 9.47[($0.3 \pm 0.2$ )\%]&&$9.44 \pm 0.02$\\
2. $ 3.4 <  \Delta R (\mu_0 \mu_1)   $ & 2.20 [($4.8 \pm 0.4 $)\%] &&2.19 [($4.3  \pm 0.4$ )\%] &&  2.18[($3.8  \pm 0.4$ )]\%&&$2.10 \pm 0.01 $\\
\hline
$\left. (S - S_0)/\sqrt{S_0}\right|_{\mathcal{L} = 3\, \rm ab^{-1}} $ & 3.8 && 3.4 &&3.0&&\\
&& && &&\\
\hline
\hline
\end{tabular}}
\caption{The cross section after each cut and its uncertainty for $M_{Z'}  = 30\, \gev$ in $pp$ collider.  }
\label{tab:30gev}
\end{table}
\end{centering}



\subsubsection{$pp \to Z \to 4\mu,\, M_{Z'} = 40\, \gev$}

The energies of the leptons from on-shell $Z'$ with mass 40 GeV is about $ 20 ~ \gev$ each.  In the signal, the energies of the two hardest leptons ($E^{\mu_0}$ and $E^{\mu_1}$) are peaked near 20 GeV (table \ref{tab:40gev}).

\begin{centering}
\begin{table}[H]
\makebox[\textwidth]{
\begin{tabular}{c c  c c c c c  | c  c  }
\hline
\hline
\multicolumn{8}{c}{Cross section [fb]}  \\
cut & & $\epsilon = 0.05 $ & & $\epsilon  = 0.01 $ && $ \epsilon  =0.005$&& SM \\
0. Basic cuts && 35.8&& 35.6 && 35.6&& $35.6\pm 0.1$\\ 
1.  $  E^{\mu_1}< 23 ~ \gev$ && 6.21 [$(1.4  \pm 0.2 )$\%] &&6.14 [$(0.3  \pm 0.2)$\%] &&6.14 [$(0.2 \pm 0.2)$\%]&& $6.13 \pm 0.01$  \\
3.  $ E^{\mu_0} < 25 ~ \gev $ && 0.090[$( 28.6   \pm 2 )$\%] &&  0.08  [$( 14.3  \pm  2   )$\%] &&  0.08 [$( 14.1  \pm 2  )$\%] && $0.07  \pm 0.002$ \\
\hline
$\left. (S - S_0)/\sqrt{S_0}\right|_{\mathcal{L} = 3\,\rm ab^{-1}} $ &&4.1  && 2.1 &&2.0 &&\\
&& && && &&\\
\hline
\hline
\end{tabular}}
\caption{The cross section after each cut and its uncertainty for $M_{Z'}  = 40\, \gev$ in $pp$ collider. }
\label{tab:40gev}
\end{table}
\end{centering} 



\section{Appendix-B}
\label{Appendix-B}

The analysis of this section is based on the study of $ pp \rightarrow \mu^+ \mu^- \slashed E_T$ for $M_{Z'} = 0.1~ \gev,$ and $0.05 ~ \gev$ and $\epsilon = 0.001$. The series of the optimal cut to search for very light $Z'$ is shown in Table \ref{tab:CutsForSig1}. The main purpose of the first three cuts is to remove the biggest background which is the $ \tau \tau$ background. Requiring the transverse between dilepton and $\slashed E_T$ to be large reduces the $\tau \tau$ background greatly, while not hurting the signal too much. High $\slashed E_T$ requirement is mainly to eliminate the effect the DY process($ p p \rightarrow \mu^+ \mu^- ~ \slashed E_T$), but it is also helpful in removing $ \tau \tau$ background. And finally after making $ \tau \tau$ background negligible by the $ \Delta \phi ( \mu_1, \slashed E_T)$ cut, we impose $ m_T(\mu^+ , \mu^-) < 50  ~ \gev$ to remove part of the di-boson background. With these cuts and $2.6 ~ \rm ab^{-1}$ of luminosity, we can get $5\sigma$ significance for these benchmark points. See table \ref{tab:2mumet} for the details of the cuts and signal to background ratio. 

\subsubsection{$pp \to 2\mu + \slashed E_T,\, M_{Z'} = 0.1\, \gev\, \rm and \, M_{Z'} = 0.05\, \gev$}

 \begin{centering}
\begin{table}[H]
\makebox[\textwidth]{
\begin{tabular}{l c  c | c c c  }
\hline
\hline
very light $Z'$, $M_{Z'} < 2 M_{\mu}$ cuts&\multicolumn{3}{c}{Cross section [fb]} & \\
\hline
Cut &   $M_{Z'} = 0.05 \gev$&   $M_{Z'} = 0.1 \gev$ & SM& $\tau \tau$ Bkg\\
&   $\& \epsilon = 0.001$+SM &   $\& \epsilon = 0.001$+SM& &  \\
\hline
$0. \left\{ \begin{matrix}p_T^{\mu_1} >~ 17 ~\gev\\
p_T^{\mu_2} >~ 8~ \gev\\
|\eta(\mu_i)| < 2.7 \\
\Delta R_{\mu \mu} > 0.4 \\
M_{\mu \mu} > 0.3 \gev
\end{matrix}\right.$  &$722.57 $ &$ 722.19 $& 722.23 & 111030   \\ 
 & & & &   \\
 1. $m_T( \mu^+\mu^-, \slashed E_T) > 100 ~\gev $&  $578.5 \pm 0.8$  &$577.7  \pm 1.3 $& $578.0 \pm  0.8 $& $984.2 \pm 11$ \\
  & & & &   \\
  2. $\slashed E_T$$  > 45 \gev$& $370.4 \pm  0.6$   &$370.6\pm 1 $&$ 307.4 \pm  0.6 $& $170.3\pm 5$ \\
  & & & &    \\
3. $\left|\Delta \phi( \mu_1, \slashed E_T)\right| < 2.5 $& $100.1 \pm  0.3$   &$100.1 \pm 0.5$&$ 99.9 \pm  0.3 $& 0  \\
  & & & &    \\
4. $ m_T(\mu^+ , \mu^-) < 50  ~ \gev$& $17.3 \pm  0.14$   &$17.5 \pm 0.26$&$ 16.9 \pm  0.1 $& $0$\\
  & & & &   \\
\hline 
   $\frac{S-S_0}{S_0 + \tau\tau}$ & $2.4\pm0.6\%$ & $2.4 \pm 1.1 \%$& & &  \\
   $\left. \frac{\sqrt{S-S_0}}{S_0 + \tau\tau}\right|_{\mathcal{L} = 2.6\, \rm ab^{-1}}$ & 5.0& 7.4 & & \\
    & & & &   \\
\hline
\hline
\end{tabular}}
\label{tab:CutsForSig1}
\caption{ The series of cuts needed in order to get $5\, \sigma$ significance for the two benchmark points $(M_{Z'}, \epsilon) = ( 0.05~ \gev, 0.001)$ and $(0.1~ \gev, 0.001)$. The luminosity needed for this benchmark point is $2.6\, \rm ab^{-1}$. }
\label{tab:2mumet}
\end{table}
\end{centering}


\section{Appendix-C}
\label{Appendix-C}


This section shows the cut flow for the cuts used in the lepton collider for the channel  $e^+ e^-  \rightarrow Z \rightarrow 4 \mu$. The center of mass in this analysis is $ \sqrt{s} = 92~ \gev$. The base cuts are exactly the same as the hadron collider analysis. We kept the masses for our this analysis the same as the LHC ($M_{Z'} = 0.5 , 10, 20 , 30 , \rm and ~ 40 ~ \gev$), but chose smaller couplings $ \epsilon = 0.01,~ 0.005,$ and $0.001$. The values in the square brackets in tables~\ref{tab:ptfiveee}-\ref{tab:40GeVee} below are $100 \times (\sigma(e^+ e^-  \rightarrow 4\mu)_{\rm SM+Z'} - \sigma(e^+ e^- \rightarrow 4\mu)_{\rm SM} )/   \sigma(e^+ e^-\rightarrow 4\mu)_{\rm SM}$, and the significance after $2.6\, \rm ab^{-1}$ is shown in the last row. 

\subsubsection{ $e^+e^- \to 4\mu, M_{Z'} = 0.5~ \gev$}

The invariant mass between two of the muons ($ \mu_1^+$ and $\mu_0^-$) can effectively discriminate signal against background that with $2.6\, \rm ab^{-1}$ we can get more than $5\,\sigma$ significance as shown in table \ref{tab:ptfiveee}. 


\begin{centering}
\begin{table}[H]
\makebox[\textwidth]{
\begin{tabular}{c c  c c c c c  c | c  }
\hline
\hline
\multicolumn{8}{c}{Cross section [fb]}  \\
cut & &  $\epsilon = 0.01 $ & & $\epsilon  = 0.005$ && $ \epsilon  =0.001$ && SM [fb]  \\
0. Basic cuts &&69.4&& 69.2 && 68.7 &&  $68.7 \pm 0.18$\\
1. $0.7~ \gev < M_{\mu_1^+ \mu_0^-} < 1.5~\gev$  &&1.92  [$(11  \pm 1)$\%] && 1.87  [$(7.7  \pm  1)$\%] && 1.87 [$(7.9 \pm 1)$\%] &&$1.73 \pm 0.03$\\
\hline
$\left. (S - S_0)/\sqrt{S_0}\right|_{\mathcal{L} = 2.6 \rm ab^{-1}} $ &&7.4  && 5.4 &&5.4&&\\
&& && && &&\\
\hline
\hline
\end{tabular}}
\caption{The cross section after each cut and its uncertainty for $M_{Z'} = 0.5 ~ \gev$ in an $e^+ e^-$ collider. }
\label{tab:ptfiveee}
\end{table}
\end{centering}



\subsubsection{ $e^+e^- \to 4\mu, M_{Z'} = 10~ \gev$}

The best cut for this $M_{Z'}$, is requiring the transverse mass of the second leading positive muon and the leading negative muon to be near $M_{Z'}$. See table \ref{tab:10GeVee} for the details of the cut.


\begin{centering}
\begin{table}[H]
\makebox[\textwidth]{
\begin{tabular}{c c  c c c c c  c | c  }
\hline
\hline
\multicolumn{8}{c}{Cross section [fb]}  \\
cut & &  $\epsilon = 0.01 $ & & $\epsilon  = 0.005$ && $ \epsilon  =0.001$ && SM [fb]  \\
0. Basic cuts && 69.1&&  68.7&& 68.7&& $68.7 \pm 0.18 $ \\
1. $9~\gev < m_T (\mu_1^+, \mu_0^-) <10~\gev$  && 1.28 [$(9.2  \pm 1.1)$\%] && 1.27 [$(7.8  \pm  1.1)$\%] && 1.26[$( 7.5\pm 1.1)$\%] &&$1.17  \pm 0.02$\\

\hline
$\left. (S - S_0)/\sqrt{S_0}\right|_{\mathcal{L} = 2.6 \rm ab^{-1}} $ &&5.2 && 4.7 &&4.2&&\\
&& && && &&\\
\hline
\hline
\end{tabular}}
\caption{The cross section after each cut and its uncertainty for $M_{Z'} = 10 ~ \gev$ in an $e^+ e^-$ collider. }
\label{tab:10GeVee}
\end{table}
\end{centering}


\subsubsection{ $e^+e^- \to 4\mu, M_{Z'} = 20~ \gev$}

Requiring  $  m_T (\mu_0^+, \mu_1^-)$ to be near $M_{Z'}$ can improve our sensitivity to $\epsilon = 0.001$ with more than $4\sigma$ significance (table \ref{tab:20GeVee}). 

\begin{centering}
\begin{table}[H]
\makebox[\textwidth]{
\begin{tabular}{c c  c c c c c  c | c  }
\hline
\hline
\multicolumn{8}{c}{Cross section [fb]}  \\
cut & &  $\epsilon = 0.01 $ & & $\epsilon  = 0.005$ && $ \epsilon  =0.001$ && SM [fb]  \\
0. Basic cuts&& 68.7&&  68.7&& 68.7&& $68.7 \pm 0.18 $ \\
1. $18~\gev < m_T (\mu_0^+, \mu_1^-) <20~\gev$  && 2.57 [$(1.8  \pm 1 )$\%] && 2.55 [$(1.7  \pm  1)$\%] && 2.53[$(1.3 \pm 1)$\%] &&$2.40  \pm 0.3$\\
\hline
$\left. (S - S_0)/\sqrt{S_0}\right|_{\mathcal{L} = 2.6 \rm ab^{-1}} $ && 5.6 && 4.9 &&4.3 &&\\
&& && && &&\\
\hline
\hline
\end{tabular}}
\caption{The cross section after each cut and its uncertainty for $M_{Z'} = 20 ~ \gev$ in an $e^+ e^-$ collider. }
\label{tab:20GeVee}
\end{table}
\end{centering}



\subsubsection{ $e^+e^- \to 4\mu, M_{Z'} = 30~ \gev$}

 A cut on $m_T(\mu_0^+ \mu_0^-) $ in the region of $M_{Z} - M_{Z'}$ is the most optimal cut for this mass. An extra cut on the separation between two leptons can further enhance our significance. See table \ref{tab:30GeVee} for more detailed information about the cuts. 


\begin{centering}
\begin{table}[H]
\makebox[\textwidth]{
\begin{tabular}{c c  c c c c c  c | c  }
\hline
\hline
\multicolumn{8}{c}{Cross section [fb]}  \\
cut & &  $\epsilon = 0.01 $ & & $\epsilon  = 0.005$ && $ \epsilon  =0.001$ && SM [fb]  \\
0. Basic cuts && 68.7&&  68.7&& 68.7&& $68.7 \pm 0.18 $ \\
1.  $59~\gev < m_T (\mu_0^+, \mu_0^-) < 60~\gev $ && 1.69   [$(5 \pm 1)$\%] && 1.68 [$(5 \pm  1)$\%] && 1.67[$(4  \pm 1)$\%] &&$1.60 \pm 0.2$\\
2. $\Delta R (\mu_2 \mu_3) <  0.25 $ &&  0.76 [$(9 \pm 2 )$\%] && 0.75 [$(8 \pm  2)$\%] && 0.75 [$(8 \pm 2)$\%] &&$ 0.69 \pm 0.01$\\
\hline
$\left. (S - S_0)/\sqrt{S_0}\right|_{\mathcal{L} = 2.6 \rm ab^{-1}} $ &&4.3 && 3.7 &&3.7 &&\\
&& && && &&\\
\hline
\hline
\end{tabular}}
\caption{The cross section after each cut and its uncertainty for $M_{Z'} = 30~ \gev$ in an $e^+ e^-$ collider.}
\label{tab:30GeVee}
\end{table}
\end{centering}



\subsubsection{ $e^+e^- \to 4\mu, M_{Z'} = 40\, \gev$}

The most efficient cut for $M_{Z'}$  is when $ m_T (\mu_0^+ \mu_0^-)$ is almost $Z$ mass. Due to phase space suppression, the rate of on-shell $Z'$ for $M_{Z'} = 40\, \gev$ is very low and thus our significance here is lower than other benchmark points as shown in table \ref{tab:40GeVee}.

\begin{centering}
\begin{table}[H]
\makebox[\textwidth]{
\begin{tabular}{c c  c c c c c  c | c  }
\hline
\hline
\multicolumn{8}{c}{Cross section [fb]}  \\
cut & &  $\epsilon = 0.01 $ & & $\epsilon  = 0.005$ && $ \epsilon  =0.001$ && SM [fb]  \\
0. Basic cuts &&  68.7&&  68.7&& 68.7&& $68.7 \pm 0.18 $ \\
1. $ 47 ~ \gev < m_T(\mu_0^+ \mu_0^-) < 49~ \gev$ && 3.01  [$(4 \pm 1)$\%] && 3.01 [$(4 \pm 1)$\%] && 3.00[$(3 \pm 1)$\%] &&$2.90 \pm 0.03$\\

\hline
$\left. (S - S_0)/\sqrt{S_0}\right|_{\mathcal{L} = 2.6 \rm ab^{-1}} $ &&3.3 && 3.3&&3.0&&\\
&& && && &&\\
\hline
\hline
\end{tabular}}
\caption{The cross section after each cut and its uncertainty for $M_{Z'} = 40~ \gev$ in $e^+ e^-$ collider.  }
\label{tab:40GeVee}
\end{table}
\end{centering}


\bibliography{mutau}

\providecommand{\href}[2]{#2}\begingroup\raggedright\begin{thebibliography}{10}

\bibitem{Schuh:2015hda}
{\bf ATLAS} Collaboration, N.~Schuh, {\it {Search for heavy resonances with the
  ATLAS detector}},  {\em EPJ Web Conf.} {\bf 95} (2015) 04061.

\bibitem{Charaf:2015jua}
{\bf CMS} Collaboration, O.~Charaf, {\it {Search for heavy resonances at CMS}},
   {\em J. Phys. Conf. Ser.} {\bf 623} (2015), no.~1 012007.

\bibitem{He:1991qd}
X.-G. He, G.~C. Joshi, H.~Lew, and R.~Volkas, {\it {Simplest Z-prime model}},
  {\em Phys.Rev.} {\bf D44} (1991) 2118--2132.

\bibitem{Eigen:2015rea}
{\bf BaBar} Collaboration, G.~Eigen, {\it {Direct Searches for New Physics
  Particles at BABAR}},  {\em J. Phys. Conf. Ser.} {\bf 631} (2015), no.~1
  012034, [\href{http://xxx.lanl.gov/abs/1503.02860}{{\tt 1503.02860}}].

\bibitem{Curtin:2014cca}
D.~Curtin, R.~Essig, S.~Gori, and J.~Shelton, {\it {Illuminating Dark Photons
  with High-Energy Colliders}},  {\em JHEP} {\bf 02} (2015) 157,
  [\href{http://xxx.lanl.gov/abs/1412.0018}{{\tt 1412.0018}}].

\bibitem{Eigen:2015sea}
{\bf BaBar} Collaboration, G.~Eigen, {\it {Recent BABAR Results}},  {\em J.
  Phys. Conf. Ser.} {\bf 631} (2015), no.~1 012033,
  [\href{http://xxx.lanl.gov/abs/1503.02867}{{\tt 1503.02867}}].

\bibitem{Essig:2013vha}
R.~Essig, J.~Mardon, M.~Papucci, T.~Volansky, and Y.-M. Zhong, {\it
  {Constraining Light Dark Matter with Low-Energy $e^+e^-$ Colliders}},  {\em
  JHEP} {\bf 11} (2013) 167, [\href{http://xxx.lanl.gov/abs/1309.5084}{{\tt
  1309.5084}}].

\bibitem{Wang:2015hdf}
B.~Wang, {\it {Searches for New Physics at the Belle II Experiment}},  in {\em
  {Meeting of the APS Division of Particles and Fields (DPF 2015) Ann Arbor,
  Michigan, USA, August 4-8, 2015}}, 2015.
\newblock \href{http://xxx.lanl.gov/abs/1511.00373}{{\tt 1511.00373}}.

\bibitem{Anastasi:2015oea}
A.~Anastasi, {\it {The Muon g-2 experiment at Fermilab}},  {\em EPJ Web Conf.}
  {\bf 96} (2015) 01002.

\bibitem{Baek:2001kca}
S.~Baek, N.~G. Deshpande, X.~G. He, and P.~Ko, {\it {Muon anomalous g-2 and
  gauged L(muon) - L(tau) models}},  {\em Phys. Rev.} {\bf D64} (2001) 055006,
  [\href{http://xxx.lanl.gov/abs/hep-ph/0104141}{{\tt hep-ph/0104141}}].

\bibitem{Ma:2001md}
E.~Ma, D.~P. Roy, and S.~Roy, {\it {Gauged L(mu) - L(tau) with large muon
  anomalous magnetic moment and the bimaximal mixing of neutrinos}},  {\em
  Phys. Lett.} {\bf B525} (2002) 101--106,
  [\href{http://xxx.lanl.gov/abs/hep-ph/0110146}{{\tt hep-ph/0110146}}].

\bibitem{Gninenko:2001hx}
S.~N. Gninenko and N.~V. Krasnikov, {\it {The Muon anomalous magnetic moment
  and a new light gauge boson}},  {\em Phys. Lett.} {\bf B513} (2001) 119,
  [\href{http://xxx.lanl.gov/abs/hep-ph/0102222}{{\tt hep-ph/0102222}}].

\bibitem{Pospelov:2008zw}
M.~Pospelov, {\it {Secluded U(1) below the weak scale}},  {\em Phys. Rev.} {\bf
  D80} (2009) 095002, [\href{http://xxx.lanl.gov/abs/0811.1030}{{\tt
  0811.1030}}].

\bibitem{Heeck:2011wj}
J.~Heeck and W.~Rodejohann, {\it {Gauged L\_mu - L\_tau Symmetry at the
  Electroweak Scale}},  {\em Phys. Rev.} {\bf D84} (2011) 075007,
  [\href{http://xxx.lanl.gov/abs/1107.5238}{{\tt 1107.5238}}].

\bibitem{Harigaya:2013twa}
K.~Harigaya, T.~Igari, M.~M. Nojiri, M.~Takeuchi, and K.~Tobe, {\it {Muon g-2
  and LHC phenomenology in the $L\_\mu-L\_\tau$ gauge symmetric model}},  {\em
  JHEP} {\bf 03} (2014) 105, [\href{http://xxx.lanl.gov/abs/1311.0870}{{\tt
  1311.0870}}].

\bibitem{Crivellin:2015lwa}
A.~Crivellin, G.~D?Ambrosio, and J.~Heeck, {\it {Addressing the LHC flavor
  anomalies with horizontal gauge symmetries}},  {\em Phys. Rev.} {\bf D91}
  (2015), no.~7 075006, [\href{http://xxx.lanl.gov/abs/1503.03477}{{\tt
  1503.03477}}].

\bibitem{Crivellin:2015mga}
A.~Crivellin, G.~D?Ambrosio, and J.~Heeck, {\it {Explaining
  $h\to\mu^\pm\tau^\mp$, $B\to K^* \mu^+\mu^-$ and $B\to K \mu^+\mu^-/B\to K
  e^+e^-$ in a two-Higgs-doublet model with gauged $L_\mu-L_\tau$}},  {\em
  Phys. Rev. Lett.} {\bf 114} (2015) 151801,
  [\href{http://xxx.lanl.gov/abs/1501.00993}{{\tt 1501.00993}}].

\bibitem{Heeck:2014qea}
J.~Heeck, M.~Holthausen, W.~Rodejohann, and Y.~Shimizu, {\it {Higgs ??? in
  Abelian and non-Abelian flavor symmetry models}},  {\em Nucl. Phys.} {\bf
  B896} (2015) 281--310, [\href{http://xxx.lanl.gov/abs/1412.3671}{{\tt
  1412.3671}}].

\bibitem{Mishra:1991ws}
{\bf CCFR} Collaboration, S.~Mishra {\em et.~al.}, {\it {Recent electroweak
  results from the CCFR Collaboration: Neutrino tridents and W - Z interference
  and the Lorentz structure of the weak current}}, .

\bibitem{Mishra:1991bv}
{\bf CCFR} Collaboration, S.~R. Mishra {\em et.~al.}, {\it {Neutrino tridents
  and W Z interference}},  {\em Phys. Rev. Lett.} {\bf 66} (1991) 3117--3120.

\bibitem{Geiregat:1990gz}
{\bf CHARM-II} Collaboration, D.~Geiregat {\em et.~al.}, {\it {First
  observation of neutrino trident production}},  {\em Phys.Lett.} {\bf B245}
  (1990) 271--275.

\bibitem{Altmannshofer:2014cfa}
W.~Altmannshofer, S.~Gori, M.~Pospelov, and I.~Yavin, {\it {Quark flavor
  transitions in $L_\mu-L_\tau$ models}},  {\em Phys.Rev.} {\bf D89} (2014)
  095033, [\href{http://xxx.lanl.gov/abs/1403.1269}{{\tt 1403.1269}}].

\bibitem{Gninenko:2014pea}
S.~N. Gninenko, N.~V. Krasnikov, and V.~A. Matveev, {\it {Muon g-2 and searches
  for a new leptophobic sub-GeV dark boson in a missing-energy experiment at
  CERN}},  {\em Phys. Rev.} {\bf D91} (2015) 095015,
  [\href{http://xxx.lanl.gov/abs/1412.1400}{{\tt 1412.1400}}].

\bibitem{delAguila:2014soa}
F.~del Aguila, M.~Chala, J.~Santiago, and Y.~Yamamoto, {\it {Collider limits on
  leptophilic interactions}},  {\em JHEP} {\bf 03} (2015) 059,
  [\href{http://xxx.lanl.gov/abs/1411.7394}{{\tt 1411.7394}}].

\bibitem{Altmannshofer:2014pba}
W.~Altmannshofer, S.~Gori, M.~Pospelov, and I.~Yavin, {\it {Neutrino Trident
  Production: A Powerful Probe of New Physics with Neutrino Beams}},  {\em
  Phys.Rev.Lett.} {\bf 113} (2014) 091801,
  [\href{http://xxx.lanl.gov/abs/1406.2332}{{\tt 1406.2332}}].

\bibitem{Beyer:1993jt}
R.~Beyer {\em et.~al.}, {\it {Neutrino electron scattering results from CHARM
  II}},  in {\em {International Europhysics Conference on High-energy Physics
  Marseille, France, July 22-28, 1993}}, 1993.

\bibitem{Smith:1992iw}
W.~H. Smith {\em et.~al.}, {\it {Neutrino scattering results from CCFR}},  {\em
  Conf. Proc.} {\bf C9207131} (1992) 487--502.

\bibitem{CMS:2012bw}
{\bf CMS} Collaboration, S.~Chatrchyan {\em et.~al.}, {\it {Observation of Z
  decays to four leptons with the CMS detector at the LHC}},  {\em JHEP} {\bf
  12} (2012) 034, [\href{http://xxx.lanl.gov/abs/1210.3844}{{\tt 1210.3844}}].

\bibitem{Aad:2014wra}
{\bf ATLAS} Collaboration, G.~Aad {\em et.~al.}, {\it {Measurements of
  Four-Lepton Production at the Z Resonance in pp Collisions at $\sqrt s=$7 and
  8 TeV with ATLAS}},  {\em Phys. Rev. Lett.} {\bf 112} (2014), no.~23 231806,
  [\href{http://xxx.lanl.gov/abs/1403.5657}{{\tt 1403.5657}}].

\bibitem{Alloul:2013bka}
A.~Alloul, N.~D. Christensen, C.~Degrande, C.~Duhr, and B.~Fuks, {\it
  {FeynRules 2.0 - A complete toolbox for tree-level phenomenology}},  {\em
  Comput.Phys.Commun.} {\bf 185} (2014) 2250--2300,
  [\href{http://xxx.lanl.gov/abs/1310.1921}{{\tt 1310.1921}}].

\bibitem{Degrande:2011ua}
C.~Degrande, C.~Duhr, B.~Fuks, D.~Grellscheid, O.~Mattelaer, {\em et.~al.},
  {\it {UFO - The Universal FeynRules Output}},  {\em Comput.Phys.Commun.} {\bf
  183} (2012) 1201--1214, [\href{http://xxx.lanl.gov/abs/1108.2040}{{\tt
  1108.2040}}].

\bibitem{Alwall:2011uj}
J.~Alwall, M.~Herquet, F.~Maltoni, O.~Mattelaer, and T.~Stelzer, {\it {MadGraph
  5 : Going Beyond}},  {\em JHEP} {\bf 1106} (2011) 128,
  [\href{http://xxx.lanl.gov/abs/1106.0522}{{\tt 1106.0522}}].

\bibitem{Watt:2012tq}
G.~Watt and R.~S. Thorne, {\it {Study of Monte Carlo approach to experimental
  uncertainty propagation with MSTW 2008 PDFs}},  {\em JHEP} {\bf 08} (2012)
  052, [\href{http://xxx.lanl.gov/abs/1205.4024}{{\tt 1205.4024}}].

\bibitem{Ball:2014uwa}
{\bf NNPDF} Collaboration, R.~D. Ball {\em et.~al.}, {\it {Parton distributions
  for the LHC Run II}},  {\em JHEP} {\bf 04} (2015) 040,
  [\href{http://xxx.lanl.gov/abs/1410.8849}{{\tt 1410.8849}}].

\bibitem{Sjostrand:2006za}
T.~Sjostrand, S.~Mrenna, and P.~Z. Skands, {\it {PYTHIA 6.4 Physics and
  Manual}},  {\em JHEP} {\bf 05} (2006) 026,
  [\href{http://xxx.lanl.gov/abs/hep-ph/0603175}{{\tt hep-ph/0603175}}].

\bibitem{Koratzinos:2013ncw}
M.~Koratzinos {\em et.~al.}, {\it {TLEP: A High-Performance Circular $e^+e^-$
  Collider to Study the Higgs Boson}},  in {\em {Proceedings, 4th International
  Particle Accelerator Conference (IPAC 2013)}}, p.~TUPME040, 2013.
\newblock \href{http://xxx.lanl.gov/abs/1305.6498}{{\tt 1305.6498}}.

\bibitem{Zimmermann:2014qxa}
F.~Zimmermann, M.~Benedikt, D.~Schulte, and J.~Wenninger, {\it {Challenges for
  Highest Energy Circular Colliders}},  in {\em {Proceedings, 5th International
  Particle Accelerator Conference (IPAC 2014)}}, p.~MOXAA01, 2014.

\bibitem{Fan:2014vta}
J.~Fan, M.~Reece, and L.-T. Wang, {\it {Possible Futures of Electroweak
  Precision: ILC, FCC-ee, and CEPC}},  {\em JHEP} {\bf 09} (2015) 196,
  [\href{http://xxx.lanl.gov/abs/1411.1054}{{\tt 1411.1054}}].

\bibitem{Ruan:2014xxa}
M.~Ruan, {\it {Higgs measurement at e+e- circular colliders}},
  \href{http://xxx.lanl.gov/abs/1411.5606}{{\tt 1411.5606}}.

\bibitem{Bai:2000pr}
{\bf BES} Collaboration, J.~Z. Bai {\em et.~al.}, {\it {First measurement of
  the branching fraction of the decay psi(2S) ---> tau+ tau-}},  {\em Phys.
  Rev.} {\bf D65} (2002) 052004,
  [\href{http://xxx.lanl.gov/abs/hep-ex/0010072}{{\tt hep-ex/0010072}}].

\end{thebibliography}\endgroup
\bibliographystyle{JHEP}

\end{document}